\documentclass[twocolumn,PRD,showpacs,nofootinbib,amsmath,amssymb,superscriptaddress]{revtex4}

\pdfoutput=1

\usepackage{graphicx}
\usepackage{dcolumn}
\usepackage{bm}
\usepackage{hyperref}
\usepackage{color}
\usepackage{rotate}
\usepackage{fancyhdr} 
\usepackage{cancel}
\usepackage{multirow}
\usepackage{ulem}

\def\be{\begin{equation}}
\def\ee{\end{equation}}
\def\bea{\begin{eqnarray}}
\def\eea{\end{eqnarray}}

\newcommand{\beq}{\begin{equation}}
\newcommand{\eeq}{\end{equation}}

\definecolor{darkred}{RGB}{175,0,0}
\definecolor{darkblue}{RGB}{14,0,185}

\newcommand{\change}[1]{{#1}}
\newcommand{\NEWchange}[1]{{#1}}

\usepackage{aas_macros,braket}
\begin{document}
\title{A User's Guide to Extracting Cosmological Information from Line-Intensity Maps}
\author{Jos\'e Luis Bernal}
\affiliation{ICC, University of Barcelona, IEEC-UB, Mart\' i i Franqu\` es 1, E08028 Barcelona, Spain}
\affiliation{Dept.\ de F\' isica Qu\` antica i Astrof\' isica, Universitat de Barcelona, Mart\' i i Franqu\` es 1, E08028 Barcelona, Spain}
\affiliation{Institut d'Estudis Espacials de Catalunya (IEEC), E08034 Barcelona, Spain}
\author{Patrick C.~Breysse}
\affiliation{Canadian Institute for Theoretical Astrophysics, University of Toronto, 60 St.\ George Street, Toronto, ON, M5S 3H8, Canada}
\author{H\'{e}ctor Gil-Mar\'{i}n}
\affiliation{ICC, University of Barcelona, IEEC-UB, Mart\' i i Franqu\` es 1, E08028 Barcelona, Spain}
\affiliation{Institut d'Estudis Espacials de Catalunya (IEEC), E08034 Barcelona, Spain}
\author{Ely D.~Kovetz}
\affiliation{Department of Physics, Ben-Gurion University, Be'er Sheva 84105, Israel}

\begin{abstract}
Line-intensity mapping (LIM) provides a promising way to probe cosmology, reionization and galaxy evolution. However, its sensitivity to  cosmology and astrophysics at the same time is also a nuisance. Here we develop a comprehensive framework for modelling the LIM power spectrum, which includes redshift space distortions and the Alcock-Paczynski effect.  We then identify and isolate degeneracies with astrophysics so that they can be marginalized over. We study the gains of using the multipole expansion of the anisotropic power spectrum, providing an accurate analytic expression for their covariance, and find a \NEWchange{$10\%$-$75\%$} increase in the precision of the baryon acoustic oscillation scale measurements when including the hexadecapole in the analysis. We  discuss different observational strategies when targeting other cosmological parameters, such as the sum of neutrino masses or primordial non-Gaussianity, finding that fewer and wider \change{redshift} bins are typically  optimal. 
Overall, our formalism facilitates an optimal extraction of cosmological constraints robust to astrophysics.

\end{abstract} 

\maketitle

\section{Introduction}\label{sec:Intro}
Line-intensity mapping (LIM)~\cite{Kovetz_IMstatus} has recently arisen as a key technique to surpass and expand  the remarkable achievements of precision observational cosmology over the past decades. 
Impressive experimental efforts concentrated mostly on two observables, namely the cosmic microwave background (CMB) and galaxy number counts, have attained percent-level measurements of the standard cosmological model, $\Lambda$-Cold Dark Matter ($\Lambda$CDM), and provided stringent constraints on possible deviations from it~\citep{Planck18_pars,alam_bossdr12}. However, the former only allows limited (due to damping on small scales) access to a brief moment in the early history of the Universe, while the latter loses effectiveness as it probes deeper into the Universe (as the discrete sources become too faint to detect).

LIM provides a way to bridge the gap between the CMB and galaxy surveys, and probe the huge swaths of the observable Universe that remain  uncharted~\cite{Karkare:2018sar,Kovetz_IM2020}. It measures the integrated emission  from either atomic or molecular spectral lines originating from all the galaxies---individually detectable or not---as well as from the diffuse intergalactic medium along the line of sight, and generates three-dimensional maps of the targeted volume that  in principle contain all the information that can be harvested from the incoming photons. 

LIM's experimental landscape is rapidly progressing. Probably the most studied line to date is the 21-cm spin-flip transition in neutral hydrogen \change{(see e.g.,~\cite{Villaescusa-Navarro_21cmingredients,Baugh_milleniumHI,Spinelli_21cm} for recent and detailed theoretical modelling)}, first detected in \change{cross-correlation with the spatial distribution of galaxies} (see Ref.~\cite{Chang_21cmdetection}). A comprehensive suite of experiments is on its way to map this line across the history of the observable Universe, with several experiments targeting the epochs of cosmic dawn and reionization~\citep{Haarlem_lofar,Tingay_MWA,hera,Koopmans_SKAlow} and others focusing on lower redshifts~\citep{Masui_GBT,CHIME_pathfinder,HIRAX,Santos_Meerkat,redbook,Battye_bingo}. 
Meanwhile, growing attention has been given to other lines. Some examples include: carbon monoxide (CO) rotational lines~\citep{Lidz2011,Pullen2013,Breysse2014,Li_CO_16,Padmanhaban_CO}, [CII]~\citep{Silva_CII,Pullen_CII,Padmanabhan2018_CII}, H$\alpha$ and H$\beta$~\citep{Gong_lines,Silva_Halpha}, oxygen lines~\citep{Gong_lines} and  Lyman-$\alpha$~\citep{Pullen2014,Silva_Lyalpha}. A few of these have already been (at least tentatively) detected at intermediate redshifts~\citep{Keating_COPSS,Keating_COPSSII,Pullen_CII,Yang2019,Croft_Lyalphadetection}.  A significant effort is now being invested in LIM experiments targeting these lines, with some instruments already observing and others to come online soon~\cite{Cleary_COMAP,TIME,Hetdex,spherex,Concerto,CCAT-prime,Starfire,aim-co,Cooray_CDIM,OST_origins}.

LIM holds unique promise for the study of both cosmology and astrophysics~\citep{Kovetz_IMstatus,Kovetz_IM2020}. 
Since the line emission  originates from halos, the intensity of the spectral lines acts as a biased tracer of the underlying density distribution (sourced by the cosmological primordial fluctuations), with the bias depending on the specific line. Meanwhile, the signal is intimately related to various astrophysical processes that take place during reionization and galaxy evolution~\citep{Breysse2016,Serra2016,Breysse2017,Wolz2017,Lidz2018,Padmanabhan2018_CII,Breysse2019}. 
A primary challenge is therefore to disentangle between the astrophysical and cosmological information contained in the maps. 

Previous works have investigated the potential of using LIM observations to study cosmology within the confines of $\Lambda$CDM and beyond it (see e.g., Refs.~\cite{Bull_21cm,Raccanelli_ska,Carucci_wdm21cm,Pourtsidou_21cmcosmo,Karkare_IMBAO,Fonseca_multitracerlines,Obuljen_21cmcrossothers,Munoz_chargedDM21cm,Chen_hiddenvalley,Moradinezhad_Dizga_COfNL1,Moradinezhad_Dizga_COfNL2,redbook,SKA_fundamental,
Heneka_21cmMG,Castorina_21cmgrowth,Foreman_lensing21cm,Schaan_IMlensing,Jalilvand_21cmgallensing}). However, existing studies are limited in the sense that they do not fully account for or properly model one or more of the following: the observable signal, observing limitations, intrument response, degeneracies between astrophysics and cosmology, or the noise covariance.  Ignoring these effects may result in a significant underestimation of the errors and bias of the best-fit parameters even when dealing merely with forecasts (see e.g., Refs.~\cite{Bellomo_Fisher,Bernal_Fisher} for some examples related to galaxy surveys). 

In this paper we aim to address this deficit by providing a general and comprehensive formalism to optimally extract robust cosmological information from the power spectrum of fluctuations in observed line-intensity maps. We focus especially on maximizing the precision of baryon acoustic oscillation (BAO) measurements, but also consider the potential of LIM to constrain extensions to $\Lambda$CDM, such as primordial non-Gaussianity or neutrinos with a sum of masses higher than $0.06\,{\rm eV}$.  
In particular, we consistently account for the following:
 \begin{itemize}
 \item {\it Degeneracies between cosmology and astrophysics.} We identify and isolate the cosmological information encoded in the  power spectrum that can be extracted given known astrophysical dependencies.
 \item {\it LIM power spectrum anisotropies.} We present a model for the  power spectrum which includes the anisotropic information from redshift-space distortions and the Alcock-Paczynski effect (see below).
  \item {\it Multipole decomposition of the  power spectrum.} We show the optimal way to account for these anisotropies, including the full covariance between different multipoles. We demonstrate generically the importance of going to higher multipoles.
 \item {\it Effects of survey volume and instrument response.}  We show how to accurately model the suppression of the power spectrum on large and small scales due to survey and beam shapes.  We discuss the effects of redshift binning and experimental optimization. 
 \end{itemize}
 In order to remain as generic as possible in our quantitative estimates, we refrain from forecasting for specific experiments and consider a generic LIM instrument which can easily be replaced with any concrete experimental setup. 
 This manuscript (hopefully) results in a useful manual for LIM enthusiasts who wish to precisely quantify the cosmological information accessible in their observed maps.

To perform the analyses this work is based on, we modify the public code \texttt{LIM}\footnote{\url{https://github.com/pcbreysse/lim}}. We  analytically compute the anisotropic LIM power spectrum and the corresponding covariance, using outputs from \texttt{CAMB}\footnote{\url{https://camb.info/}}~\citep{CAMB} to extract the matter power spectrum and other cosmological quantities, and \texttt{Pylians}\footnote{\url{https://github.com/franciscovillaescusa/Pylians}}~\citep{VillaescusaNavarro_pylians} to obtain the halo mass function and halo bias. We plan to release an updated version of the LIM package upon publication.

Our calculations are based on several standard assumptions and parameter choices. 
Throughout this work we adopt the halo mass function and halo bias presented in Ref.~\cite{Tinker_hmf2010}.  We assume, unless otherwise stated, the Planck baseline $\Lambda$CDM model and take the best-fit parameter values from the combined analysis of the full Planck 2018 and galaxy BAO measurements~\citep{Planck18_pars,alam_bossdr12}:  baryon and dark matter physical densities at $z=0$, $\Omega_{\rm b}h^2 = 0.0224$ and $\Omega_{\rm cdm}h^2 = 0.1193$, respectively; spectral index $n_s=0.967$; amplitude $\log \left(A_s10^{10}\right) = 3.047$ of the primordial power spectrum of scalar perturbations at the pivot scale; Hubble constant $H_0 = 67.67$ km/s/Mpc; a sum of neutrino masses $\sum m_\nu=0.06$ eV; and Gaussian primordial perturbations, i.e., $f_{\rm NL}=0$.

This paper is structured as follows. In Section~\ref{sec:ModellingPK} we review the basics of LIM power spectrum modelling. We highlight the degeneracies between astrophysics and cosmology and provide a formulation which clearly delineates them. Properties of the measured power spectrum are derived in Section~\ref{sec:MeasuredPK}, including the Legendre multipole expansion, and the effects of the window function and the  covariance matrix. A generalized LIM experiment is introduced in Section~\ref{sec:Forecasting}, along with the prescription for our Fisher-based forecasts. In  Section~\ref{sec:ExtractingCosmology} we investigate the extraction of cosmological information from the LIM power spectrum, demonstrating the importance of higher multipoles when measuring BAO and RSD, and describing strategies to improve cosmological constraints, including redshift binning and experimental optimization. We defer to an Appendix the adaptation of our methodology to cross-correlations between different lines or with galaxy surveys (Appendix~\ref{sec:Cross}), as well as the discussion of how to fold in the instrument response and the signal suppression due to the finite volume surveyed, and the effects of foregrounds and line interlopers (Appendix~\ref{sec:Error}). We present our conclusions in Section~\ref{sec:Conclusions}.

\section{Modelling the IM Power Spectrum}\label{sec:ModellingPK}

\subsection{From line luminosity to the power spectrum}
The brightness temperature $T$ and specific intensity $I$ of a given radiation source are equivalent quantities that depend on the expected luminosity density $\rho_L$ of a spectral line with rest frame frequency $\nu$ at redshift $z$:
\begin{equation}
\begin{split}
& T(z) = \frac{c^3(1+z)^2}{8\pi k_B\nu^3H(z)} \rho_L(z), \\
& I(z) = \frac{c}{4\pi\nu H(z)}\rho_L(z),
\end{split}
\label{eq:TofL}
\end{equation}
where $c$ is the speed of light, $k_B$ is the Boltzmann constant and $H$ is the Hubble parameter \cite{Lidz2011}. Depending on the frequency band of the experiment, either $T$ or $I$ are conventionally used.  Hereinafter, we will use $T$ in order to homogenize the nomenclature, but all the expressions are equally valid if intensity is used instead. Assuming a known relation, $L(M,z)$, between the luminosity of the spectral line and the mass $M$ of the host halo, the expected mean luminosity density  can be computed using the halo mass function ${\rm d}n/{\rm d}M(z)$:
\begin{equation}
\langle \rho_L\rangle(z) = \int {\rm d}ML(M,z)\frac{{\rm d}n}{{\rm d}M}(M,z).
\label{eq:rhoL}
\end{equation}

The main statistic we will focus on in this work is the LIM power spectrum, which is given by the Fourier transform of the two-point correlation function of the perturbations of the brightness temperature, denoted by $\delta T\equiv T-\langle T\rangle$. Since spectral lines are sourced in halos, $\delta T$ can be used to trace the halo distribution and, equivalently, the underlying matter density perturbations. At linear order, matter density and brightness temperature perturbations are related by an effective linear bias, which is given in terms of the halo bias $b_{\rm h}$ by
\begin{equation}
b(z) = \frac{\int{\rm d}ML(M,z)b_{\rm h}(M,z)\frac{{\rm d}n}{{\rm d}M}(M,z)}{\int {\rm d}ML(M,z)\frac{{\rm d}n}{{\rm d}M}(M,z)}.
\label{eq:bias_IM}
\end{equation}

In practice, $T$ maps are obtained in redshift space, where the observed position along the line of sight is disturbed with respect to real space due to peculiar velocities, producing  the so-called redshift-space distortions (RSD). RSD introduce anisotropies in the a-priori isotropic real-space power spectrum. The relation between real and redshift space perturbations ($\delta^r$ and $\delta^s$, respectively) can be approximated by a linear factor, such that $\delta^s=F_{\rm RSD}\delta^r$, with $F_{\rm RSD}$ in Fourier space given by:
\begin{equation}
F_{\rm RSD}(k,\mu,z) = \left(1+ \frac{f(z)}{b(z)}\mu^2\right)\frac{1}{1+0.5\left(k\mu\sigma_{\rm FoG}\right)^2},
\label{eq:RSD}
\end{equation}
where $f(z)$ is the logarithmic derivative of the growth factor (also known as the growth rate), $k$ is the \change{modulus} of the Fourier mode, and $\mu= \hat{k}\cdot\hat{k}_\parallel$ is the cosine of the angle between the mode vector $\vec{k}$ and the line-of-sight component, $k_\parallel$ (i.e.\ $\mu\in \left[-1,1\right]$). At linear scales, coherent peculiar velocities boost the power spectrum in redshift space through the Kaiser effect (first term in Eq.~\eqref{eq:RSD}) \cite{Kaiser1987}. In turn, small-scale velocities suppress the clustering on small scales, an effect known as the fingers of God (second term in Eq.~\eqref{eq:RSD}). We use a Lorentzian damping factor whose scale-dependence is driven by the parameter $\sigma_{\rm FoG}$, the value of which is related to the halo velocity dispersion\footnote{The fingers-of-God damping  can be also modelled with a Gaussian function, yielding similar results. We have checked that the conclusions of this work do not depend on this choice.}. We assume a fiducial value of $7\,{\rm Mpc}$. We refer the interested reader to e.g., Ref.~\cite{Hamilton_RSDreview}, for a review on RSD discussing these contributions. 

On top of all these effects, there is a scale-independent shot noise contribution to the LIM power spectrum stemming from the discreteness of the source population, which would be present even in the absence of clustering. Thus we can express the anisotropic LIM power spectrum as the sum of clustering and shot noise contributions:
\begin{equation}
\begin{split}
& P(k,\mu,z) = P_{\rm clust}(k,\mu,z) + P_{\rm shot}(z);\\ 
& P_{\rm clust}(k,\mu,z) = \langle T\rangle^2(z)  b^2(z) F_{\rm RSD}^2(k,\mu,z)P_{\rm m}(k,z); \\
& P_{\rm shot}(z)  = \left(\frac{c^3(1+z)^2}{8\pi k_B\nu^3H(z)}\right)^2 \int dM L^2(M,z)\frac{{\rm d}n}{{\rm d}M},
\end{split}
\label{eq:Pk}
\end{equation}
where $P_{\rm m}$ is the matter power spectrum.

We note that the above expressions assume that each halo contains only a single point-source line emitter at its center.  In many cases, there will be wider-scale emission from diffuse gas within a halo, or there will be satellite galaxies which contribute additional intensity to that of the central emitter.  Either of these will lead to an additional ``one-halo" contribution to the clustering term which traces the halo density profile.  We direct the interested reader to Ref. \cite{Wolz2019} for a discussion of one-halo effects in intensity mapping surveys. We also leave for future work the effect of anisotropic halo assembly bias~\cite{Obuljen:2019ukz}.

\subsection{Dependence on astrophysics and cosmology}\label{sec:astro}
The emission of any spectral line is intimately related to astrophysical processes. Therefore, the observed density field is strongly dependent on astrophysics through the line luminosity function. \change{ Fortunately, according to Eq.~\eqref{eq:Pk}, the relation between astrophysics and the LIM power spectrum at linear order is limited to the amplitude of the clustering and shot noise contributions\footnote{A detailed study of the nonlinearity and scale dependence of the astrophysical terms in the LIM power spectrum lies beyond the scope of this work.}. Then, astrophysical information enters the power spectrum in three different ways.  The overall amplitude is determined by the mean intensity of the line $\langle T\rangle$ and the bias $b$.  For optically thin lines like 21 cm, $\langle T\rangle$  is proportional to the mean density of the emitting gas ($\Omega_HI$ in the case of 21 cm).  On the other hand, the mean intensity of the optically thick CO line can be related to the density $\Omega_{\rm H_2}$ through the $\alpha_{\rm CO}$ parameter (see e.g.,~\cite{Bolatto_COH2}). The effective linear bias is the halo bias weighted by the luminosity function (see Eq.~\ref{eq:bias_IM}); since the most massive halos are the most strongly biased, the determination of the bias is dominated by the mass of the brightest halos. Finally, line  luminosity functions tend to cut off at some characteristic luminosity $L_\star$, so that, for a typical model, the shot noise $P_{\rm shot}$ is dominated by these $L_\star$ galaxies.}   This means that the only accessible astrophysical information present in the 
 LIM power spectrum is  contained in the first two moments of $L(M)$, \change{as well as its first moment weighted by the halo bias}\footnote{Alternative probes such as the voxel intensity distribution~\cite{Breysse_VID} are more suitable to infer $L(M)$, see Section~\ref{sec:Conclusions}.}.

Since $\delta T$ traces matter overdensities, LIM observations also carry cosmological information. This information is mostly encoded in $P_{\rm m}$, $F_{\rm RSD}$ and the BAO. For example, as seen in Eq.~(\ref{eq:RSD}), $F_{\rm RSD}$ depends on the growth factor $f$, and so can be efficiently used to constrain deviations from general relativity. 

In order to measure the power spectrum, redshifts need to be transformed into distances. This procedure introduces further anisotropies in the power spectrum if the assumed cosmology does not match the actual one. This is known as the Alcock-Paczynski effect~\cite{Alcock_paczynski}. Radial and transverse distances are distorted in different ways, which can be modelled by introducing rescaling parameters to redefine distances:
\begin{equation}
\alpha_\perp = \frac{D_A(z)/r_{\rm s}}{\left( D_A(z)/r_{\rm s}\right)^{\rm fid}},\qquad \alpha_\parallel = \frac{\left(H(z)r_{\rm s}\right)^{\rm fid}}{H(z)r_{\rm s}},
\label{eq:scaling}
\end{equation}
where $D_A$ is the angular diameter distance, $r_{\rm s}$ is the sound horizon at radiation drag, and the superscript `${\rm fid}$' denotes the corresponding values in the assumed (fiducial) cosmology. Due to this distortion, the \textit{true} wave numbers are related to the \textit{measured} ones in the transverse and line-of-sight directions by $k^{\rm true}_\perp=k^{\rm meas}_\perp/\alpha_\perp$ and $k^{\rm true}_\parallel=k^{\rm meas}_\parallel/\alpha_\parallel$, respectively. These  relations can then be expressed in terms of $k$ and $\mu$~\citep{Ballinger_scaling96}:
\begin{equation}
\begin{split}
& k^{\rm true} = \frac{k^{\rm meas}}{\alpha_\perp}\left[1+\left(\mu^{\rm meas}\right)^2\left(F_{\rm AP}^{-2}-1\right)	\right]^{1/2}, \\
& \mu^{\rm true} = \frac{\mu^{\rm meas}}{F_{\rm AP}}\left[1+\left(\mu^{\rm meas}\right)^2\left(F_{\rm AP}^{-2}-1\right)	\right]^{-1/2},
\end{split}
\label{eq:scaling_kmu}
\end{equation}
where $F_{\rm AP} \!\equiv\! \alpha_\parallel/\alpha_\perp$. To correct for the modification of the volumes, the power spectrum is then multiplied by the factor: $H(z)/H^{\rm fid}(z)\times \left( D_A^{\rm fid}(z)/D_A(z)\right)^2$.  BAO provide useful cosmic rulers, as they allow high precision measurements of $\alpha_\perp$ and $\alpha_\parallel$, which can effectively constrain the expansion history of the Universe, see Ref.~\cite{Bernal_IM_letter}. Moreover, the Alcock-Paczynski effect has been proposed as a method to identify and remove line interlopers~\cite{Lidz_APforeground}.

Other cosmological dependences in Eq.~\eqref{eq:Pk} arise through the halo bias and the halo mass function, and also due to the halo velocity dispersion (via $\sigma_{\rm FoG}$). However, the cosmological information encoded in these quantities will be degenerate with the highly-uncertain $L(M)$ modelling. Therefore,  in our analysis below we will mostly focus on BAO and RSD.
 
\subsection{Explicitly accounting for the degeneracies}\label{sec:cosmo}

As clearly evident, the quantities appearing in Eq.~\eqref{eq:Pk} present strong degeneracies. Therefore, after identifying where the cosmological information is encoded, it is useful to reparametrize Eq.~\eqref{eq:Pk} to minimize the degeneracies. 

The overall shape of $P_{\rm m}$ is not very sensitive to small variations of the cosmological parameters of $\Lambda$CDM. Thus, it is preferable to employ a template for $P_{\rm m}(k,z)$, computed based on the fiducial cosmology (parametrizing its amplitude with $\sigma_8$, the root mean square of the density fluctuations within $8\, h{\rm Mpc}^{-1}$), and then measure the terms that relate it to the LIM power spectrum. 

Examining Eq.~\eqref{eq:Pk}, we find that $\sigma_8$, $b$ and $\langle T\rangle$ are degenerate within  a given single redshift, as well as $\sigma_8$, $f\mu^2$, and $\langle T\rangle$. These degeneracies can be broken with an external prior on $\langle T\rangle$, or using tomography and interpreting the measurements under a given cosmological model, taking advantage of their different  redshift evolutions. 

Other ways to break these degeneracies, at least partially, include the combination of the LIM power spectrum and  higher-order statistics (see e.g., Ref.~\cite{GilMarin_bispectrunNbody} in the case of halo number counts, and Ref.~\cite{Sarkar_21cmBispec} for the 21cm LIM case); multi-tracer techniques with several spectral lines and galaxy surveys~\cite{Obuljen_21cmcrossothers}; or the exploitation of the mildly non-linear regime of the LIM power spectrum by modelling the halo clustering with Lagrangian perturbation theory~\citep{Castorina_21cmgrowth}.

Taking all this into account, we group all degenerate parameters and reparameterize the LIM power spectrum from Eq.~\eqref{eq:Pk} as:
\begin{equation}
P(k,\mu) = \left(\frac{\langle T\rangle b\sigma_8  + \langle T\rangle f\sigma_8 \mu^2 }{1+0.5\left( k\mu\sigma_{\rm FoG}  \right)^2}\right)^2\frac{P_m(k,\vec{\varsigma})}{\sigma_8^2} + P_{\rm shot},
\label{eq:paramPk}
\end{equation}
where all quantities depend on $z$. From Eq.~\eqref{eq:paramPk}, the set of parameter combinations that can be directly measured from the LIM power spectrum at each independent redshift bin and observed patch of sky is:
\begin{equation}
\vec{\theta} = \lbrace  \alpha_{\perp}, \alpha_\parallel, \langle T\rangle f\sigma_8, \langle T\rangle b\sigma_8, \sigma_{\rm FoG}, P_{\rm shot},\vec{\varsigma} \rbrace.
\label{eq:parameters}
\end{equation}
In $\vec{\varsigma}$ we group all parameters that modify the template of the power spectrum used in the analysis, such as any of the standard $\Lambda$CDM parameters, or extensions like primordial non-Gaussianity or neutrino masses.

As  explained above, out of all the parameter combinations included in Eq.~\eqref{eq:parameters}, only $\alpha_\perp$, $\alpha_\parallel$ and $\langle T\rangle f\sigma_8$ contain useful cosmological information (modulo $\langle T\rangle$, in the case of $\langle T\rangle f\sigma_8$).  
Given the difficulty in tracing cosmological information hidden in the other parameter combinations, we prefer to be conservative and consider them simply as nuisance parameters in our analysis. \change{Out of these nuisance parameters, those related by the astrophysical processes triggering the emission (i.e., effectively, by $L(M)$), may be correlated. Modelling this correlation would open up the possibility to impose a correlated prior on these quantities, which would tighten the final parameter inference. Nonetheless, given the uncertainties regarding $L(M)$, we prefer to remain as agnostic as possible and assume all nuisance parameters to be uncorrelated.}

Lastly, note that we choose this parameterization in order to separate $b$ and $f$. However, the LIM power spectrum can be cast in  different ways in order to target other parameter combinations. For example, if a measurement of $\beta=f/b$ is wanted, one could use:
\begin{equation}
P(k,\mu) = \left(\frac{\langle T\rangle b\sigma_8\left(1  + \beta \mu^2\right) }{1+0.5\left( k\mu\sigma_{\rm FoG}  \right)^2}\right)^2\frac{P_{\rm m}(k)}{\sigma_8^2} + P_{\rm shot}.
\end{equation}
In this case, the measured quantities would be $\beta$ and $ \langle T\rangle b\sigma_8$, instead of $ \langle T\rangle f\sigma_8$ and $ \langle T\rangle b\sigma_8$.  

\section{Measuring the IM Power Spectrum}\label{sec:MeasuredPK}

\change{In this section we describe the impact of the experiment on the measurement of the signal, both sourced by the instrument response due to limitations in resolution and volume surveyed, and by the instrumental noise. In addition, we describe here the multipole expansion of the LIM power spectrum. We depict a general framework for single-dish experiments, and give guidelines for its application to interferometers, under simplifying approximations (it is straightforward to include in our formalism more detailed descriptions of the impact of using interferometers; for brevity and clarity, we leave that to other works). }

\subsection{The window function and multipole expansion}

LIM experiments have a limited resolution and probe a finite volume, which limits the minimum and maximum accessible scales, respectively. This effectively renders the observed brightness temperature fluctuations smoothed with respect to the true ones. This smoothing can be modelled by convolving the true $T$ map with window functions describing the instrument response and the effects due to surveying a finite volume.  This would yield an observer-space power spectrum, given by:
\begin{equation}
\begin{split}
\tilde{P}(k,\mu,z)& =  W(k,\mu,z)P(k,\mu,z) = \\
& = W_{\rm vol}(k,\mu,z)W_{\rm res}(k,\mu,z)P(k,\mu,z),
\end{split}
\label{eq:obsPk}
\end{equation}
where $W_{\rm vol}$ and $W_{\rm res}$ respectively are the survey-area and instrument-response window functions in Fourier space. 


The line-of-sight resolution of a LIM instrument is given by the width of the frequency channels. The resolution in the plane of the sky depends on whether the experiment uses only the auto-correlation of each of its antennas (i.e., a single-dish approach) or employs interferometric techniques. In the former case, the resolution is determined by the antenna's beam profile, while in the latter, by the largest baseline of the interferometer, $D_{\rm max}$. The characteristic resolution limits in the radial and transverse directions are given by:
\begin{equation}
\begin{split}
 \sigma_\parallel = &\frac{c(1+z)\delta\nu}{H(z)\nu_{\rm obs}};\\
  \sigma_\perp^{\rm dish} = \chi(z)\sigma_{\rm beam}, \quad & \quad \sigma_\perp^{\rm interf} = \frac{c\chi(z)}{\nu_{\rm obs}D_{\rm max}},
  \end{split}
 \label{eq:resolution}
 \end{equation} 
where $\delta\nu$ is the frequency-channel width, $\nu_{\rm obs}$ is the observed frequency, $\chi(z)$ is the radial comoving distance,  and the width of the beam profile is given by $\sigma_{\rm beam}=\theta_{\rm FWHM}/\sqrt{8\log 2}$ (where $\theta_{\rm FWHM}$ is its full width at half maximum). Then, the resolution window function $W_{\rm res}$ that models the instrument response in Fourier space can be computed as~\citep{Li_CO_16}:
 \begin{equation}
 W_{\rm res}(k,\mu) = \exp\left\lbrace -k^2\left[\sigma_\perp^2(1-\mu^2)+\sigma_\parallel^2\mu^2    \right]\right\rbrace.
 \label{eq:Wk_res}
 \end{equation} 
The beam profile, field of view, and frequency band window might not be Gaussian; for instance, the frequency channels are likely to be discrete bins (which would correspond to a top-hat window in real space). In practice, the exact instrument response will be accurately characterized by each experiment. Investigation of more realistic window functions is beyond the scope of this work.

On the other hand, at a given redshift, observing a patch of the sky with solid angle $\Omega_{\rm field}$ over a frequency band $\Delta\nu$ corresponds to a surveyed volume $V_{\rm field}$, which in the absence of complex observation masks is given by:
\begin{equation}
V_{\rm field} = \left[ \chi^2(z)\Omega_{\rm field}  \right]\left[ \frac{c(1+z)^2\Delta\nu}{H(z)\nu} \right].
\label{eq:Vfield}
\end{equation}  
The two factors in Eq.~\eqref{eq:Vfield} are the transverse area and the length of the radial side of the volume observed (i.e., $V_{\rm field}\sim  L_\perp^2L_\parallel$). These are the largest scales that can be probed by a single-dish like experiment, hence we define $k_\parallel^{\rm min, dish}\equiv 2\pi/L_\parallel$ and $k_\perp^{\rm min, dish}\equiv 2\pi/L_\perp$. While an interferometer shares this limitation in the radial direction, the largest scales that an interferometer can measure are determined by the shortest baseline $D_{\rm min}$, so that $k_\perp^{\rm min, interf}\equiv 2\pi\nu_{\rm obs}D_{\rm min}/(c\chi(z))$. Note that $k_\perp^{\rm min, interf}$ will always be ultimately limited by $L_\perp$. LIM power spectrum measurements beyond these scales will be suppressed by the lack of observable modes.

To account for the loss of modes, we define a volume window function $W_{\rm vol}$, which in Fourier space is given by:
\begin{equation}
\begin{split}
W_{\rm vol}(k,\mu) = & \left(1-\exp\left\lbrace -\left(\frac{k}{k^{\rm min}_\perp}\right)^2\left(1-\mu^2 \right) \right\rbrace \right)\times \\
& \times \left(1-\exp\left\lbrace -\left(\frac{k}{k^{\rm min}_\parallel}\right)^2\mu^2 \right\rbrace \right).
\end{split}
\label{eq:Wk_vol}
\end{equation}

We use a smoothed window in order to extend the loss of modes to smaller scales than those corresponding to $k_{\perp,\parallel}^{\rm min}$. This mimics the effect of having residuals present in the map after the removal of foregrounds (and line-interlopers), polluting the signal, since large scales are the ones more likely to be affected\footnote{One could assume a top-hat window in real space to perform a more aggressive analysis. Complex observational masks may entail more complicated $W_{\rm vol}$, but these effects are very dependent on the particular circumstances of a given experimental setup.}. 

Though we have worked up to this point in $(k,\mu)$ coordinates, we must point out that while the Fourier transform of the observed map is performed using Cartesian coordinates,  the line of sight changes with different pointings on the sky. It is therefore not parallel to any Cartesian axis. Therefore, it is not possible to obtain a well defined $\mu$ from the observations, meaning that we cannot directly measure $\tilde{P}(k,\mu)$. Nevertheless, it is possible to directly measure the {\it multipoles} of the anisotropic power spectrum using, e.g., the Yamamoto estimator~\cite{YamamotoEstimator}. 

Accounting for all the effects described in this section, the multipole expansion of the observed LIM power spectrum is given by:
\begin{equation}
\begin{split}
\tilde{P}_{\ell}(k^{\rm meas}) = & \frac{H(z)}{H^{\rm fid}(z)}
\left(\frac{D_A^{\rm fid}(z)}{D_A(z)}\right)^2\frac{2\ell+1}{2}  \times \\ 
& \times \int_{-1}^{1}{\rm d}\mu^{\rm meas} \tilde{P}(k^{\rm true},\mu^{\rm true})\mathcal{L}_{\ell}(\mu^{\rm meas}),
\end{split}
\label{eq:multipole_scale}
\end{equation}
where $\mathcal{L}_{\ell}$ is the Legendre polynomial of degree $\ell$. In our analysis below, we will calculate the integral in Eq.~\eqref{eq:multipole_scale} numerically. Analytic results (neglecting the fingers-of-God contribution to the RSD and not considering $W_{\rm vol}$) can be found in Ref.~\cite{Chung_IMPkvoxel}. We note that there is a non-vanishing contribution from the shot noise of the LIM power spectrum to  multipoles higher than the monopole, due to the effect of the window function.

Most previous works do account for the smearing of the brightness temperature maps, but choose to include this effect in the covariance of the LIM power spectrum rather than in the observed signal. This is equivalent to deconvolving the observed LIM power spectrum by $W(k,\mu)$. As we show in Appendix~\ref{sec:Error}, this approach may result in a reduction of the significance of the measured LIM power spectrum multipoles when their covariance is properly modelled\footnote{The contribution to the covariance of the monopole is often modelled incorrectly, see Appendix~\ref{sec:Error}.}, unless special treatment is applied to both signal and  errors. We explore this avenue and derive a nearly optimal estimator of the multipoles of the true unsmoothed LIM power spectrum. However, given the complexity of applying this approach to observations, we advocate for modelling the smoothed LIM power spectrum rather than deconvolving the observed one.

\subsection{The power spectrum covariance matrix}\label{sec:covariance}

There are three contributions to the LIM power spectrum covariance: (i) sample variance; (ii) instrumental noise; and (iii) residual contamination after removal of foregrounds and line interlopers. Since the magnitude of the effect of foregrounds and line interlopers greatly depends on the spectral line and redshift of interest, we consider only the first two contributions to the covariance and assume that the signal has been cleaned such that the effects included in $W_{\rm vol}$ are sufficient. We refer the reader interested in foreground and line-interloper removal strategies to, e.g., Refs.~\cite{Oh2003,Wang2006,Liu2011,Lidz_APforeground,Cheng_foregroundsAP,Sun_foregrounds,Breysse_foregrounds}. We also assume a Gaussian covariance without mode coupling.

The instrumental noise, set by the practical limitations of the experiment, introduces an artificial floor brightness temperature in the observed map. The noise power spectrum for a given single-dish like experiment or an interferometer is given by (see e.g., Ref.~\cite{Bull_21cm}):
 \begin{equation}
\begin{split} 
P_n^{\rm dish} = &  \frac{T_{\rm sys}^2V_{\rm field}}{\Delta\nu t_{\rm obs} N_{\rm pol}N_{\rm feeds}N_{\rm ant}}; \\
 P_n^{\rm interf} =  & \frac{T_{\rm sys}^2V_{\rm field}\Omega_{\rm FOV}}{\Delta\nu t_{\rm obs} N_{\rm pol}N_{\rm feeds} n_s},
 \end{split}
\label{eq:Pn}
 \end{equation}
 where $t_{\rm obs}$ is the total observing time, $N_{\rm feeds}$ is the number of detectors in each antenna, each of them able to measure $N_{\rm pol}=1,2$ polarizations, $T_{\rm sys}$ is the system temperature, $\Omega_{\rm FOV}=c^2/(\nu_{\rm obs}D_{\rm dish})^2$ is the field of view of an antenna, and $n_s$ is the average number density of baselines in the visibility space. Assuming a constant number density of baselines, the latter is given by~\cite{Bull_21cm}:
 \begin{equation}
 \NEWchange{n_s = \frac{c^2N_{\rm ant}(N_{\rm ant}-1)}{2\pi\nu_{\rm obs}^2(D_{\rm max}^2-D_{\rm min}^2)}.}
  \end{equation}
 
Although the anisotropic power spectrum may  not be measurable, it is useful to define the covariance per $k$ and $\mu$ bin:
\begin{equation}
\begin{split}
& \tilde{\sigma}^2(k_i,k_j,\mu)  = \frac{1}{\sqrt{N_{\rm modes}(k_i,\mu)}}\left(\tilde{P}(k_i,\mu) +P_n \right)\times\\
& \times\frac{1}{\sqrt{N_{\rm modes}(k_j,\mu)}}\left(\tilde{P}(k_j,\mu) +P_n \right)\delta_{ij}^K,
\end{split}
\label{eq:sigma2}
\end{equation}
where  $\delta^K$ is the Kronecker delta (introduced because we neglect mode coupling). In Eq.~\eqref{eq:sigma2}, the first term in the parentheses corresponds to the sample variance, and $N_{\rm modes}$ denotes the number of modes per bin in $k$ and $\mu$ in the observed field:
\begin{equation}
N_{\rm modes}(k,\mu) = \frac{k^2\Delta k\Delta\mu}{8\pi^2}V_{\rm field},
\label{eq:Nmodes}
\end{equation}
with $\Delta k$ and $\Delta\mu$ referring to the width of the $k$ and $\mu$ bins, respectively.
 
The covariance matrix of the LIM power spectrum multipoles is comprised of the sub-covariance matrices of each multipole, and those between different multipoles. The sub-covariance matrix for multipoles $\ell$ and $\ell^\prime$ is given by:
\begin{equation}
\begin{split}
\tilde{\mathcal{C}}_{\ell\ell^\prime}(k_i,k_j) = & \frac{(2\ell +1)(2\ell^\prime+1)}{2}\times
\\
& \times\int_{-1}^1{\rm d}\mu \tilde{\sigma}^2(k_i,k_j,\mu)\mathcal{L}_\ell(\mu)\mathcal{L}_{\ell^\prime}(\mu),
\end{split}
\label{eq:covariance}
\end{equation}
where the Delta function we assumed in Eq. (\ref{eq:sigma2}) makes each sub-covariance matrix diagonal. However, modelling the non-zero covariance between multipoles is \change{essential} for an unbiased analysis. We refer the interested reader to Ref.~\cite{Grieb_covmat} for a thorough analytic derivation of the covariance of the multipoles of the galaxy power spectrum  under the Gaussian assumption.

\section{Forecasting for IM Experiments}\label{sec:Forecasting}

\subsection{Parameterizing a generic IM experiment}\label{sec:experiment}

In order to illustrate the potential of our methodology, it is useful to consider a concrete example for a LIM experiment --- without loss of generality --- so that we can compute the window functions and the covariance. Without specifying the targeted emission line, we choose as our strawperson LIM experiment an ambitious single dish-like instrument with a total frequency band $\Delta\nu/\nu~\approx~0.2$. We split the corresponding volume into several redshift bins to study how the signal --- and the extraction of cosmological information --- changes with redshift. In our analysis below we use five non-overlapping, independent redshift bins centered at $z=\lbrace 2.7,4.0,5.3,6.6,7.9  \rbrace$, such that $\log_{10}\left[\Delta(1+z)\right] = \log_{10}\left[\Delta(\nu/\nu_{\rm obs})\right] = 0.1 $. 

In order to compute the signal, we need to assume some numbers for the astrophysical quantities that are present in Eq.~\eqref{eq:Pk}. Strictly as an example, we will assume a CO intensity mapping signal described by the model from Ref.~\cite{Li_CO_16}.  In the redshift bins defined above, this model gives mean brightness temperature \NEWchange{$\langle T\rangle = \lbrace 4.3,5.4,6.5,6.8,6.1\rbrace$}, luminosity-averaged bias \NEWchange{$b=\lbrace 1.4,1.9,2.3,2.8,3.5  \rbrace$}, and shot noise contribution to the LIM power spectrum \NEWchange{$P_{\rm shot}=\lbrace 1462, 1459, 995,564, 317 \rbrace$} (Mpc/$h$)$^3\mu$K$^2$.  Note that we are neglecting the evolution of these quantities within each bin, simply using the value at the central redshift each time.

As an analogy with the futuristic experiment envisioned in Ref.~\cite{Bernal_IM_letter}, we consider an array of single dish antennas with total $N_{\rm pol}N_{\rm feeds}N_{\rm ant}t_{\rm obs}/T_{\rm sys}^2 = 10500$~h/K$^2$ and $18900$ h/K$^2$ for the first two redshift bins, respectively, and $25\times 10^3$ h/K$^2$ for the last three redshift bins. The value of $N_{\rm pol}N_{\rm feeds}N_{\rm ant}t_{\rm obs}/T_{\rm sys}^2$ changes with redshift because we assume that $T_{\rm sys}$ depends on $\nu_{\rm obs}$ until it saturates: $T_{\rm sys} = {\rm max}\left[20,\,\nu_{\rm obs}\left({\rm K}/{\rm GHz}\right)\right]$ (see e.g.,~\cite{Prestage_GBT,Murphy_nextVLA}). Furthermore, we assume a spectral resolution $\nu_{\rm obs}/\delta\nu=\lbrace 15450,11500,9150,7600,6500\rbrace$, and $\theta_{\rm FWHM}=4$ arcmin. Taking into account these experiment specifications, we choose $\Omega_{\rm field} = 1000$ deg$^2$ to maximize the significance of the measurement around the scales of the BAO. For this experiment, we would have $\sigma_\perp<\sigma_\parallel$ and $L_\perp > L_\parallel$.

We emphasize that our choice of this particular CO model is only an example. Our formalism is general and applicable to any line and experiment, and the quantities we compute here can easily be calculated for any other model the reader might have in mind.

\begin{figure*}
\centering
\includegraphics[width=0.8\textwidth]{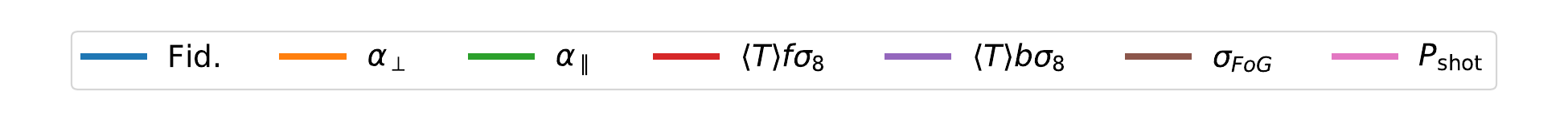}
\includegraphics[width=\textwidth]{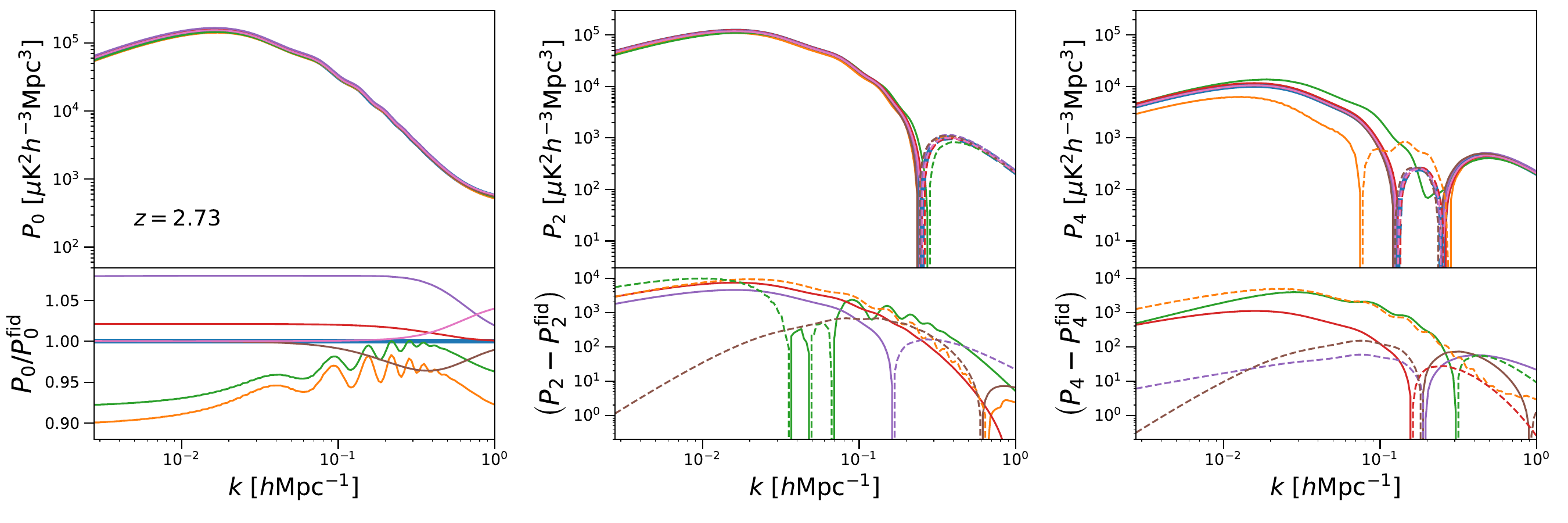}
\includegraphics[width=\textwidth]{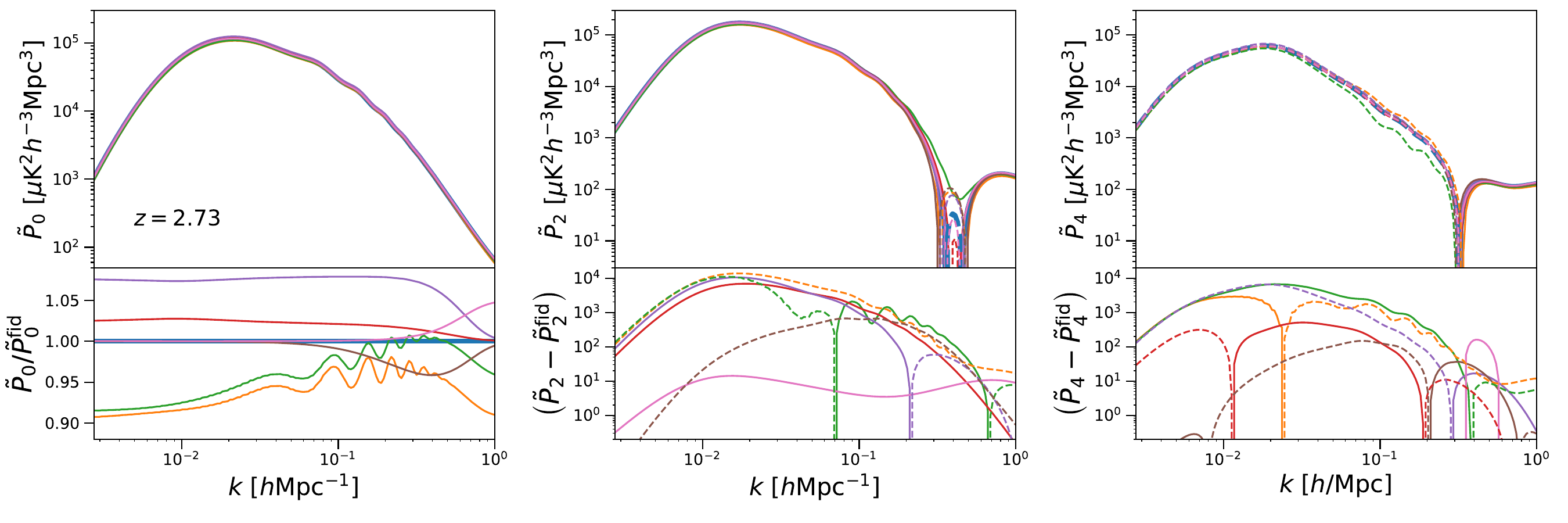}
\caption{Comparison of the LIM power spectrum multipoles at $z=2.73$  for the fiducial set of parameters (blue) and the cases where each of the  parameters of  Eq.~\eqref{eq:parameters}  is increased by $5\%$ (except for $\vec{\varsigma}$). The top panels show the true power spectrum, while the bottom panels show the smoothed power spectrum as measured by the generalized experiment considered. From left to right, each column corresponds to the monopole, quadrupole and hexadecapole, respectively. The lower part of each panel shows the ratio or difference (when the multipoles cross zero) between the fiducial power spectrum and the ones with the varied parameters. Dashed lines denote negative values. Note the effect of the window function on all multipoles, especially the quadrupole and hexadecapole: both the fiducial power spectra and their parameter dependence are considerably affected by the smoothing of the map.}
\label{fig:cosmovar}
\end{figure*}

In Fig.~\ref{fig:cosmovar} we demonstrate the effect of  the LIM power spectrum smoothing, discussed in Section~\ref{sec:MeasuredPK}, for the specific case of the first redshift bin of our strawperson experiment, comparing $P_\ell$ and $\tilde{P}_\ell$. While for the monopole, $W$ just suppresses the power spectrum at high and low $k$, the effect on the quadrupole, and especially on the hexadecapole, is much more significant. The effect largely depends on the ratio between $\sigma_\perp$ and $\sigma_\parallel$, and between $k_\perp^{\rm min}$ and $k_\parallel^{\rm min}$. Therefore, the effect of $W$ on the  power spectrum is of course strongly dependent on the experiment.  This was similarly demonstrated in Ref.~\cite{Chung_IMPkvoxel}, though the survey-volume window effect was not included there.

We also show how the LIM power spectrum multipoles change when each of the parameters of Eq.~\eqref{eq:parameters} is $5\%$ larger than the fiducial values. We can see that the derivatives of the quadrupole and hexadecapole with respect to the parameters considerably change whether we consider $P$ or $\tilde{P}$. This figure provides an intuition of the dependence of the power spectrum on each of the considered free parameters, both for the true and the measured LIM power spectra, and may guide the design of future experiments depending on the parameter targeted. 

\subsection{Likelihood and Fisher Matrix}
If we construct the vectors $\vec{\Theta}(k) = \left[\tilde{P}_0(k),\tilde{P}_2(k),...\right]$, and $\vec{\Theta} = \left[\tilde{P}_0(k_0), \tilde{P}_0(k_1) ,..., \tilde{P}_2(k_0), \tilde{P}_2(k_1), ...\right]$, we can compute S/N$(k)$, the total signal-to-noise ratio per $k$ bin, as well as the total S/N summed over all bins as:
 \begin{equation}
 \begin{split}
\left[{\rm S/N}\right(k)]^2 & =  \vec{\Theta}^T(k)\tilde{\mathcal{C}}^{-1}(k)\vec{\Theta}(k), \\ 
\left[{\rm S/N}\right]^2  & = \vec{\Theta}^T\tilde{\mathcal{C}}^{-1}\vec{\Theta}, 
\end{split}
\label{eq:SN}
\end{equation}
where the superscript $^T$ denotes the transpose operator and $\tilde{\mathcal{C}}(k_i)$ is a $N_\ell\times N_\ell$ matrix (with $N_\ell$ being the number of multipoles included in the analysis) made up of the corresponding elements of $\mathcal{C}_{\ell\ell^\prime}(k_i,k_i)$. Similarly, we can compute the $\chi^2$ of the LIM power spectrum multipoles for a given experiment, using:
\begin{equation}
\chi^2 = \Delta\vec{\Theta}^T\tilde{\mathcal{C}}^{-1}\Delta\vec{\Theta},
\label{eq:chi2}
\end{equation}
where $\Delta\vec{\Theta}$ is the difference between the model prediction and the actual measurements.

We will use the Fisher matrix formalism to forecast constraints from our generalized LIM experiment~\citep{Fisher:1935,Tegmark_fisher97}. The Fisher matrix is the average of  the second partial derivatives of the logarithm of the likelihood, $\log\mathcal{L}$, around the best fit (or assumed fiducial model). 
 Eq.~\eqref{eq:chi2} can be adapted to form the Fisher matrix by replacing $\Delta\vec{\Theta}$ by the corresponding derivatives.  The Fisher matrix element corresponding to the parameters $\vartheta_a$ and $\vartheta_b$ is:
\begin{equation}
F_{\vartheta_a\vartheta_b} = \left\langle \frac{\partial^2\log\mathcal{L}}{\partial\vartheta_a\partial\vartheta_b}  \right\rangle = 
\left(\frac{\partial\vec{\Theta}^T}{\partial\vartheta_a}\mathcal{\tilde{C}}^{-1}\frac{\partial\vec{\Theta}}{\partial\vartheta_b}\right).
\label{eq:Fisher}
\end{equation}

\section{Extracting Cosmology from LIM}\label{sec:ExtractingCosmology}

Based on the previous sections, we are now equipped to calculate quantitative estimates of  the potential of LIM experiments to set robust cosmological constraints. This is a good way to evaluate the methodology presented above. 
We will use our generic LIM experiment and forecast  constraints on $\alpha_\perp$, $\alpha_\parallel$ and $\langle T\rangle f\sigma_8$, as well as on the sum of neutrino masses and primordial non-Gaussianity.

\subsection{The importance of going to higher multipoles}

In order to exploit the anisotropic BAO, it is necessary to measure at least the monopole and quadrupole of the LIM power spectrum, Eq.~\eqref{eq:multipole_scale}. This is equivalent to constraining $D_A/r_{\rm s}$ and $Hr_{\rm s}$, rather than a combination of them if only the isotropic signal is measured, with the obvious benefits that it entails~\cite{Padmanabhan_anisoBAO}. However, while measuring the monopole and quadrupole is relatively easy, given  their reasonably high S/N, detecting the hexadecapole is challenging, since  the signal is typically well below the noise. The situation is even more pessimistic for $\ell>4$.

 Neglecting the fingers-of-God effect and $W_{\rm vol}$, Ref.~\cite{Chung_IMPkvoxel} finds that including the hexadecapole ($\ell = 4$) does not significantly improve the estimated constraints on the astrophysical parameters.  Nevertheless, focusing on cosmology and taking into account that the Alcock-Paczynski effect is anisotropic, we find that including the hexadecapole might be useful. Even if the hexadecapole is not detectable, it might help to break degeneracies between the BAO and RSD parameters, since its amplitude is mostly independent of the luminosity-averaged bias, but it does depend greatly on the $\alpha_\perp$ and $\alpha_\parallel$ configuration.  For instance, including the hexadecapole in the analysis of the quasar power spectrum in eBOSS resulted in smaller degeneracies  between $\alpha_\parallel$ and $\alpha_\perp$, and between $\alpha_\parallel$ and $f\sigma_8$, compared with the case without including the hexadecapole~\citep{GilMarin_qsoeBOSS}.
  
Moreover, the hexadecapole can be measured and estimated at the same time as the lower order multipoles, without requiring additional computing time. However, robust measurements of the hexadecapole  impose stronger requirements on experiments, given its anisotropy. In order to assess the benefits of adding the hexadecapole to the LIM power spectrum analysis, we compare the projected results with and without its inclusion.

\begin{figure}
\centering
\includegraphics[width=\linewidth]{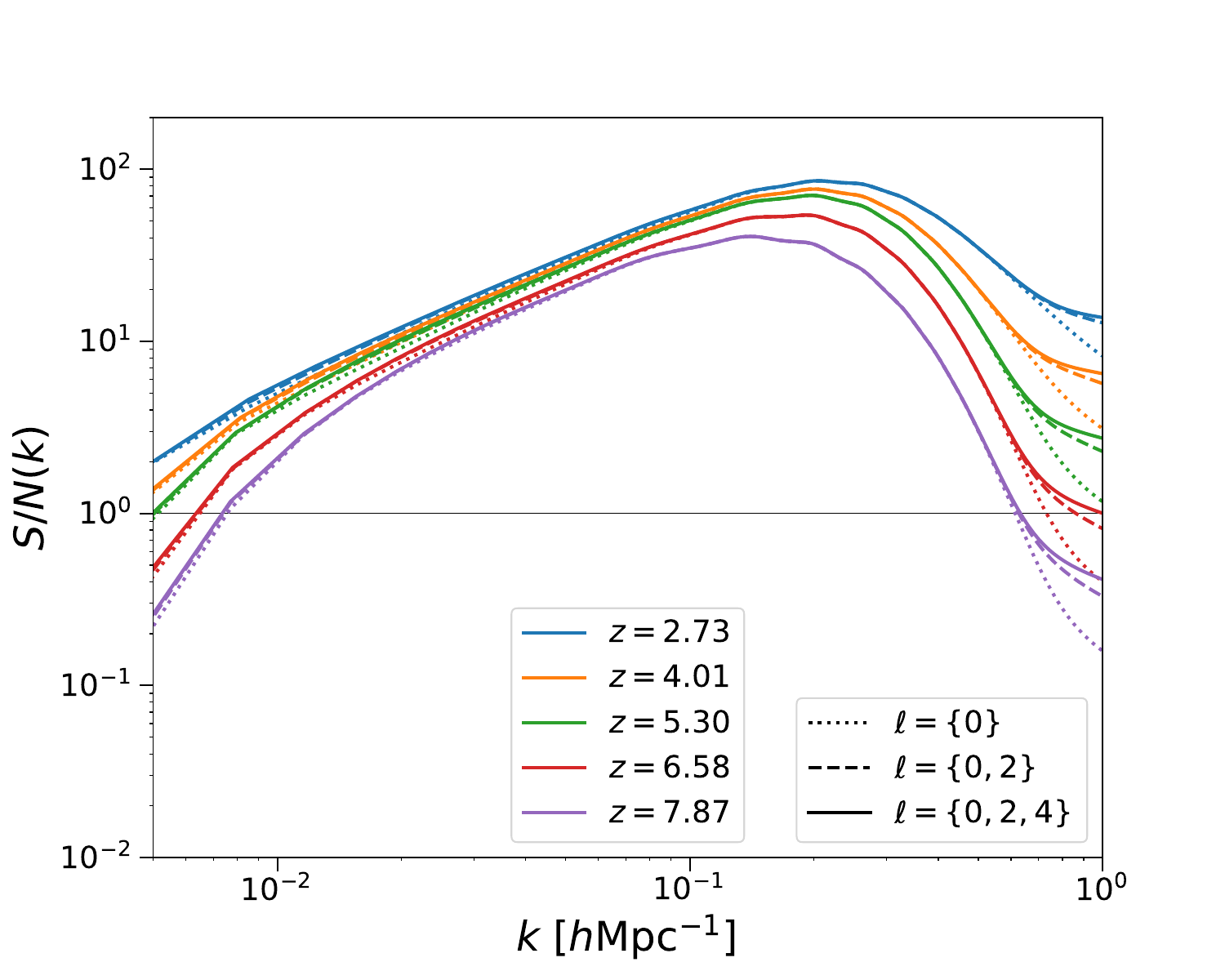}
\caption{ Signal-to-noise ratio per $k$ bin  of the measured LIM power spectrum for each redshift bin of the generic  experiment considered in this work (color coded) including only the monopole (dotted lines), the monopole and the quadrupole (dashed lines) and adding also the hexadecapole (solid lines). }
\label{fig:SNRk}
\end{figure}

We first evaluate the straightforward gain of including more multipoles than just the monopole: we compare the S/N$(k)$ of the LIM power spectrum using only the monopole and then consecutively adding the quadrupole and hexadecapole. The results given \change{by} our  strawperson experiment in each of the five redshift bins are shown in Fig.~\ref{fig:SNRk}. We can see that  adding the quadrupole or the hexadecapole does not significantly increase the total S/N per $k$ bin. Note that this does not mean that the S/N of $\tilde{P}_2$ or $\tilde{P}_4$ is  low, since the total S/N is not the sum of the individual significances of each multipole; the covariance between multipoles must be included  (see Eq.~\eqref{eq:SN}). 

 It is also evident that the signal-to-noise decreases with redshift. This is because the amplitude of the LIM power spectrum decreases with redshift faster than the noise level over the range considered here. In addition, the window function suppresses the LIM power spectrum at wider $k$ ranges.   While the behavior  in Fig.~\ref{fig:SNRk} depends on the specific experiment due to the covariance and the anisotropy of $W$, the qualitative result applies generally.

\begin{figure}
\centering
\includegraphics[width=\linewidth]{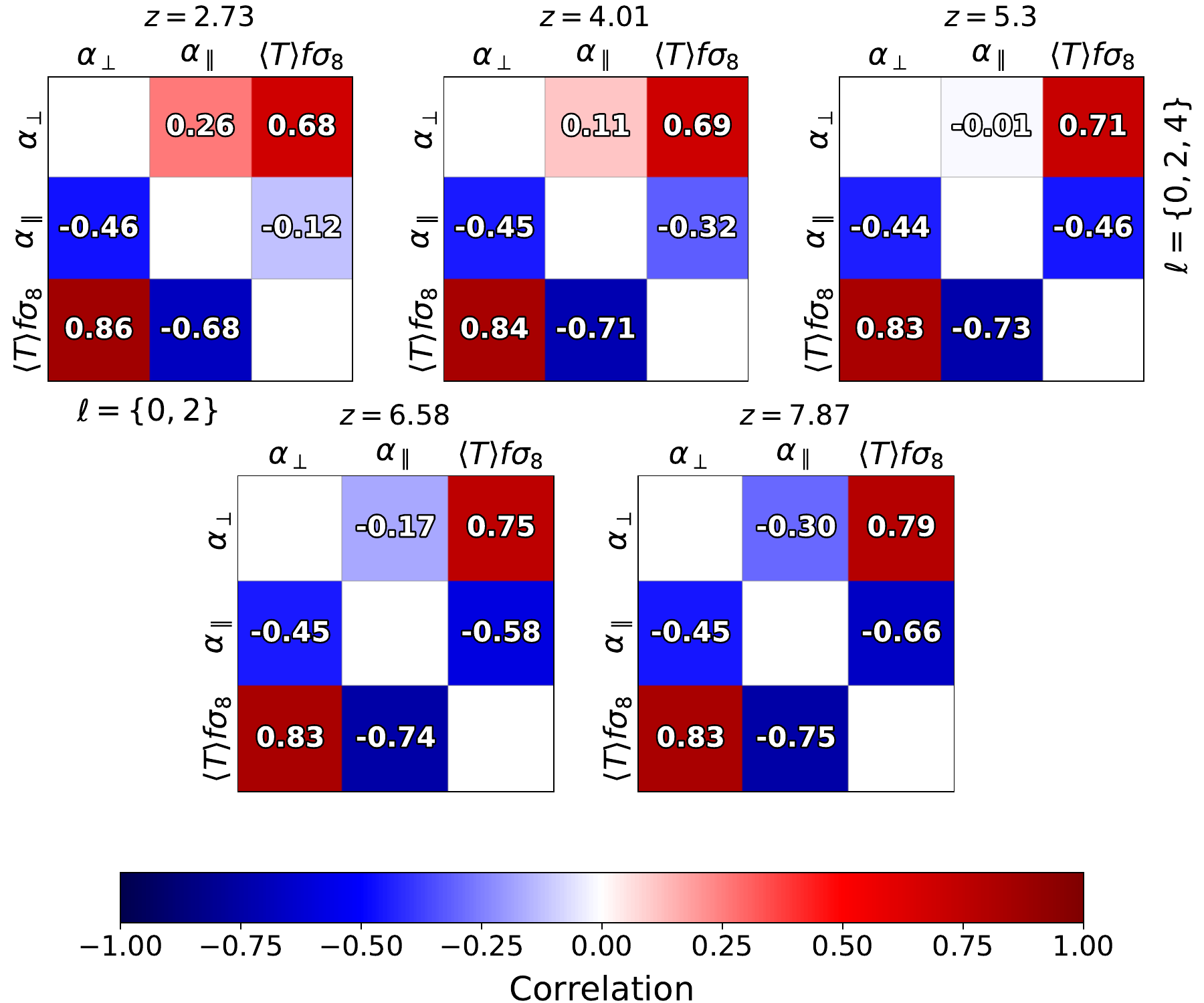}
\caption{ Correlation matrices of the parameters $\alpha_\perp$, $\alpha_\parallel$ and $\langle T\rangle f\sigma_8$, marginalized over the nuisance parameters, for each redshift bin, calculated for our generalized  experiment. The correlations in the lower triangular matrix correspond to the case where the monopole and quadrupole are included. In the upper triangular matrix the hexadecapole is included as well.}
\label{fig:correlation}
\end{figure}

Nonetheless, adding multipoles beyond the monopole helps to break degeneracies between parameters. The correlation matrices for each redshift bin after marginalizing over nuisance parameters  are shown in Fig.~\ref{fig:correlation}. We compare the results with and without  the hexadecapole. This figure shows a reduction of the correlation between the  parameters when the hexadecapole is included with respect to the case in which it is not, \change{even when the S/N of the hexadecapole is very small. Somewhat reduced degeneracy  also arises in case of a non-detection of the hexadecapole, since this is very sensitive to $\alpha_\perp$, $\alpha_\parallel$ and $f\sigma_8$, as shown in Fig.~\ref{fig:cosmovar}}. Note that this reduction is smaller for larger redshifts (where the global S/N, and especially that of the hexadecapole, is indeed smaller). 

\begin{table}[]
\vspace{0.2cm}
\centering
\resizebox{\columnwidth}{!}{%
\begin{tabular}{|c||c|c||c|c||c|c|}
\hline
 \multirow{2}{*}{$z$} & \multicolumn{2}{c||}{$\sigma_{\rm rel}\left(\alpha_\perp\right) (\%)$} & \multicolumn{2}{c||}{$\sigma_{\rm rel}\left(\alpha_\parallel\right)(\%)$} & \multicolumn{2}{c|}{$\sigma_{\rm rel}\left(\langle T \rangle f\sigma_8\right)(\%)$} \\ \cline{2-7} 
  & $\ell\leq 2$ & $\ell\leq 4$ & $\ell\leq 2$ & $\ell\leq 4$ & $\ell\leq 2 $ & $\ell\leq 4$ \\ \hline\hline
 2.73 & 1.0 & 0.7 & 1.4 & 0.8 & 3.5 & 2.0 \\ \hline
 4.01 & 0.8 & 0.6 & 1.2 & 0.8 & 3.8 & 2.4 \\ \hline
 5.30 & 0.8 & 0.6 & 1.2 & 0.8 & 4.2 & 3.0 \\ \hline
 6.58 & 1.0 & 0.9 & 1.5 & 1.2 & 5.4 & 4.9 \\ \hline
 7.87 & 1.4 & 1.3 & 2.0 & 1.7 & 5.7 & 4.9 \\ \hline
\end{tabular}
}
\caption{Forecasted 68\% confidence-level marginalized relative constraints using our generalized experiment on the BAO parameters and $\langle T \rangle f\sigma_8$, expressed as percentages. We compare results for only the monopole and quadrupole to the case of adding the hexadecapole, leading to marked improvement.}
\label{tab:lcdm_const}
\end{table}

\begin{figure}
\centering
\includegraphics[width=\linewidth]{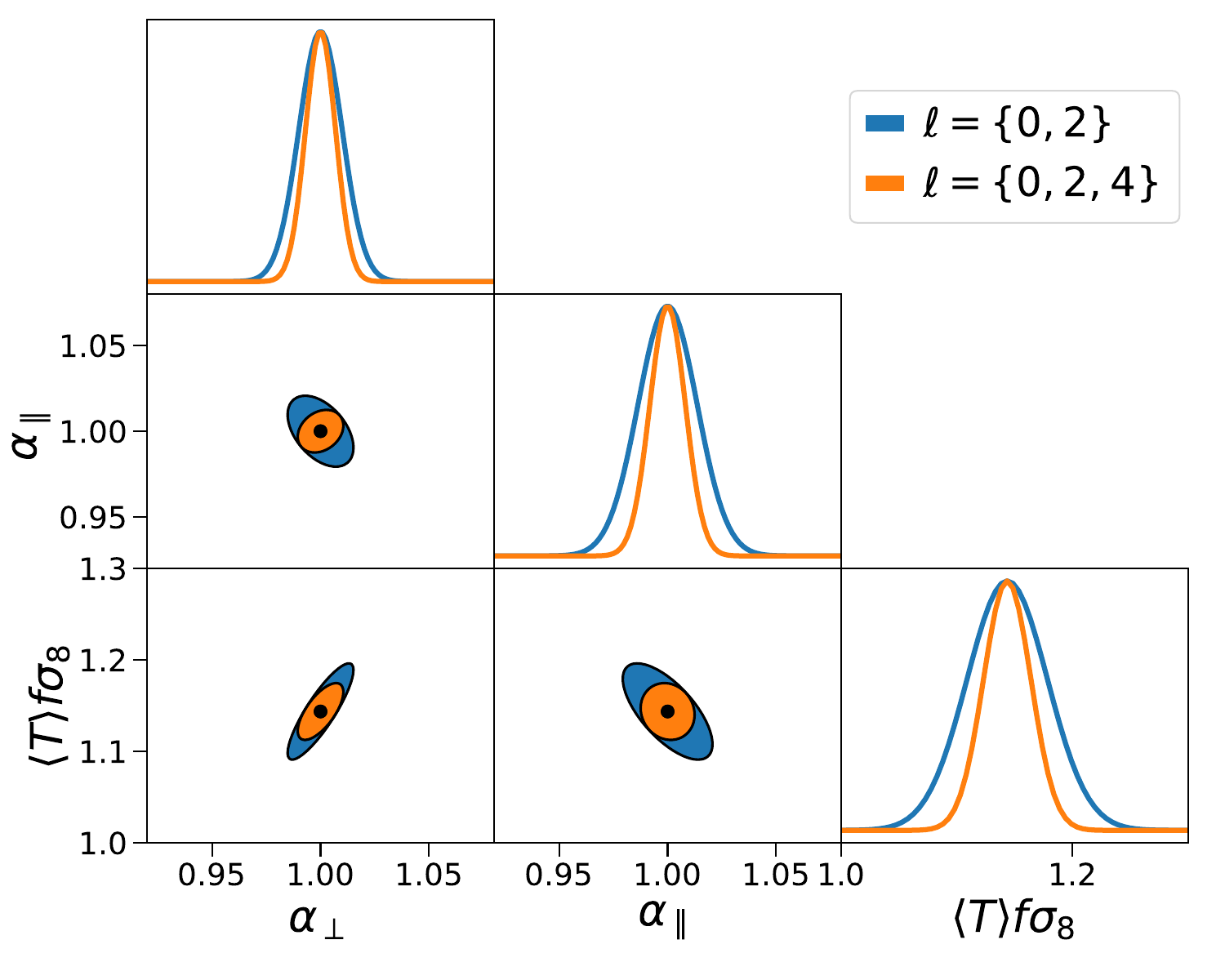}
\caption{\change{Forecasted 68\% confidence-level marginalized 2D constraints for the pair combinations $\alpha_\perp$, $\alpha_\parallel$ and $\langle T\rangle f\sigma_8$, and their corresponding marginalized one-dimensional distributions. We show results using the monopole and quadrupole (blue), and when including the hexadecapole as well (orange).}}
\label{fig:corner}
\end{figure}

Lower absolute correlations have a positive impact on the final marginalized constraints. Table~\ref{tab:lcdm_const} reports the forecasted 68\% confidence-level marginalized  precision of the measurements of the BAO rescaling parameters and the parameter combination $\langle T\rangle f\sigma_8$, for each of the redshift bins of our general experiment. The fiducial values of $\alpha_\perp$ and $\alpha_\parallel$ are 1 for all redshifts, the corresponding fiducial values of \NEWchange{$\langle T\rangle f\sigma_8$ are $\left\lbrace 1.14,\, 1.11,\, 1.07,\, 0.93,\, 0.71\right\rbrace\, \mu$K} for increasing $z$ (assuming the $\langle T\rangle$ values reported in Section~\ref{sec:experiment}). \change{The improvement of the marginalized constraints as a result of the reduced correlation between the parameters, in the case of the first redshift bin, can be seen in Fig.~\ref{fig:corner}.}

The worsening of the forecasted constraints with redshift is expected, given the decreasing S/N (as shown in Fig.~\ref{fig:SNRk}). Interestingly, the marginalized constraints on the BAO rescaling parameters are between  11\% and 62\% stronger when including the hexadecapole. The  improvement obtained by including the hexadecapole decreases with redshift, as expected from the reduction of the difference between correlations with and without the hexadexapole shown in Fig.~\ref{fig:correlation}. Although measuring even higher multipoles might  yield further gain, the improvement would be marginal, and the level of observational systematics would be too high to consider these measurements reliable. This is why we limit our study to $\ell \leq 4$.

\subsection{Redshift binning and experimental optimization}\label{sec:optimization}

When aiming to measure BAO and RSD, it is preferable to bin in redshift as much as possible, so that the evolution of the expansion of the Universe and growth of structure is better constrained. However, if the target is a parameter that does not change with redshift, wider redshift bins are more optimal, since that reduces the covariance of the LIM power spectrum (see Section~\ref{sec:covariance}). Therefore, the  analysis should always be adapted to the target parameter. \change{For instance, using the LIM angular power spectrum with infinitely-thin redshift bins, accounting for all the cross-correlations between different redshifts, should in principle return all the encoded information in the LIM fluctuations. This approach is mainly constrained by performance limitations, although it also has its drawbacks, since it includes many more nuisance parameters.  Moreover, the angular power spectrum is far less sensitive to the BAO than the anisotropic power spectrum in $k$-space. Therefore, if measuring the expansion history of the Universe is the goal of the observations, the latter strategy should be chosen. }
We illustrate this problematic by forecasting constraints on the sum $\sum m_\nu$ of the neutrino masses (when allowed to be larger than $0.06\,{\rm eV}$), and on deviations from primordial Gaussian perturbations. 

Any imprint caused by primordial deviations from Gaussian initial conditions would have been preserved in the ultra-large scales of the matter power spectrum, which have remained outside  the horizon since inflation. In the local limit, primordial non-Gaussianity can be parametrized with $f_{\rm NL}$ as the  amplitude of the local quadratic contribution of a single Gaussian random field $\phi$ to the Bardeen potential $\Phi$  \footnote{Here we assume the convention of large scale structure rather than the one for CMB ($f_{\rm NL}^{\rm LSS}\approx 1.3 f_{\rm NL}^{\rm CMB}$ \citep{Xia_PNG}).}:
\begin{equation}
\Phi(x) = \phi(x) + f_{\rm NL}\left(\phi^2(x)-\langle\phi^2\rangle\right).
\label{eq:bardeen}
\end{equation} 
The skewness introduced in the density probability distribution by local primordial non-Gaussianity increases the number of massive objects, hence introducing a scale dependence on the halo bias~\citep{Matarrese_png00,Dalal_png07,Matarrese_png08,Desjacques_png}. Denoting the Gaussian halo bias with $b^{\rm G}_h$, the total halo bias appearing in Eq.~\eqref{eq:bias_IM} is given by $b_h(k,z) = b^{\rm G}_h(z)+\Delta b_h(k,z)$, where:
\begin{equation}
\Delta b_h(k,z) =  \left[b^{\rm G}_h(z) - 1\right]f_{\rm NL}\delta_{\rm ec}\frac{3\Omega_m H_0^2}{c^2k^2T_m(k)D(z)}.
\label{eq:bias_PNG}
\end{equation}
Here $\Omega_m$ is the matter density parameter at $z=0$, $\delta_{\rm ec}=1.68$ is the critical value of the matter overdensity for ellipsoidal collapse, $D(z)$ is the linear growth factor (normalized to 1 at $z=0$) and $T_m(k)$ is the matter transfer function (which is approximately $\sim 1$ at large scales). 

The extensions we consider to $\Lambda$CDM therefore have $\sum m_\nu$ and $f_{\rm NL}$ as extra parameters, respectively, which we include in $\vec{\varsigma}$ in our Eq.~\eqref{eq:parameters}, since they modify the power spectrum template. 
Naturally, varying either $\sum m_\nu$ or $f_{\rm NL}$ would also slightly modify the halo mass function and therefore alter the LIM power spectrum via modifications of $\langle T\rangle$, $P_{\rm shot}$ and $b$. We do not model this dependence, but our findings should not be affected by this: the halo mass function (minimally) affects the amplitudes of the $P_{\rm clust}$ and  $P_{\rm shot}$ terms, which we marginalize over. 

Moreover, we should note that more massive neutrinos induce a slight scale dependence on the halo bias that can be efficiently removed if the power spectrum of baryons and cold dark matter (instead of the power spectrum of all matter, including neutrinos) is used (see e.g.,~\cite{Raccanelli_neutrino_worry,Vagnozzi_neutrino_uncorrected,Valcin_beHappy}). Since we do not aim here for detailed constraints, but for an illustration of the survey optimization, we ignore these effects and leave their study to future work.

We compare the performance using the redshift binning proposed in Section~\ref{sec:experiment} with those of a single redshift bin centered at $z=3.80$ and covering the whole frequency band of the experiment. In order to maximize the coverage at large scales, we use logarithmic binning in $k$ in both cases. All measured quantities present in Eq.~\eqref{eq:parameters}, except for $\vec{\varsigma}$ in some cases (e.g.,\ $\sum m_\nu$ or $f_{\rm NL}$), are different in each redshift bin. Therefore, the final marginalized constraints on $\vec{\varsigma}$  obtained using several redshift bins is given by the result of the combination of each individual marginalized constraint obtained from each redshift bin:
\begin{equation}
\sigma_{\vartheta_\varsigma} = \left\lbrace\sum_z \left[\left(F_z^{-1}\right)_{\vartheta_\varsigma\vartheta_\varsigma}\right]^{-1}\right\rbrace^{-1/2},
\end{equation}
where $F_z$ is the Fisher matrix of a given redshift bin.

\begin{figure*}
\centering
\includegraphics[width=0.48\textwidth]{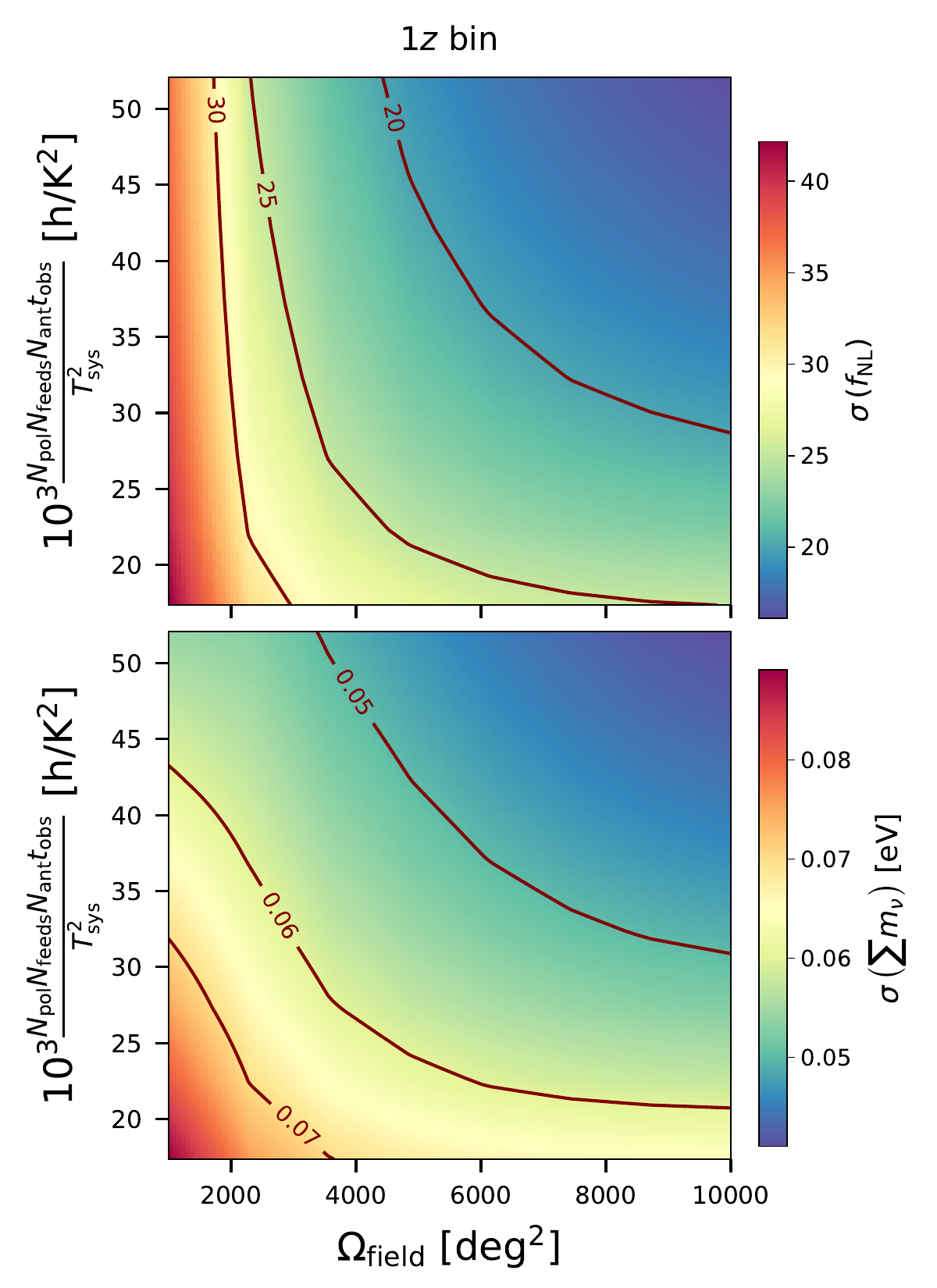}
\includegraphics[width=0.48\textwidth]{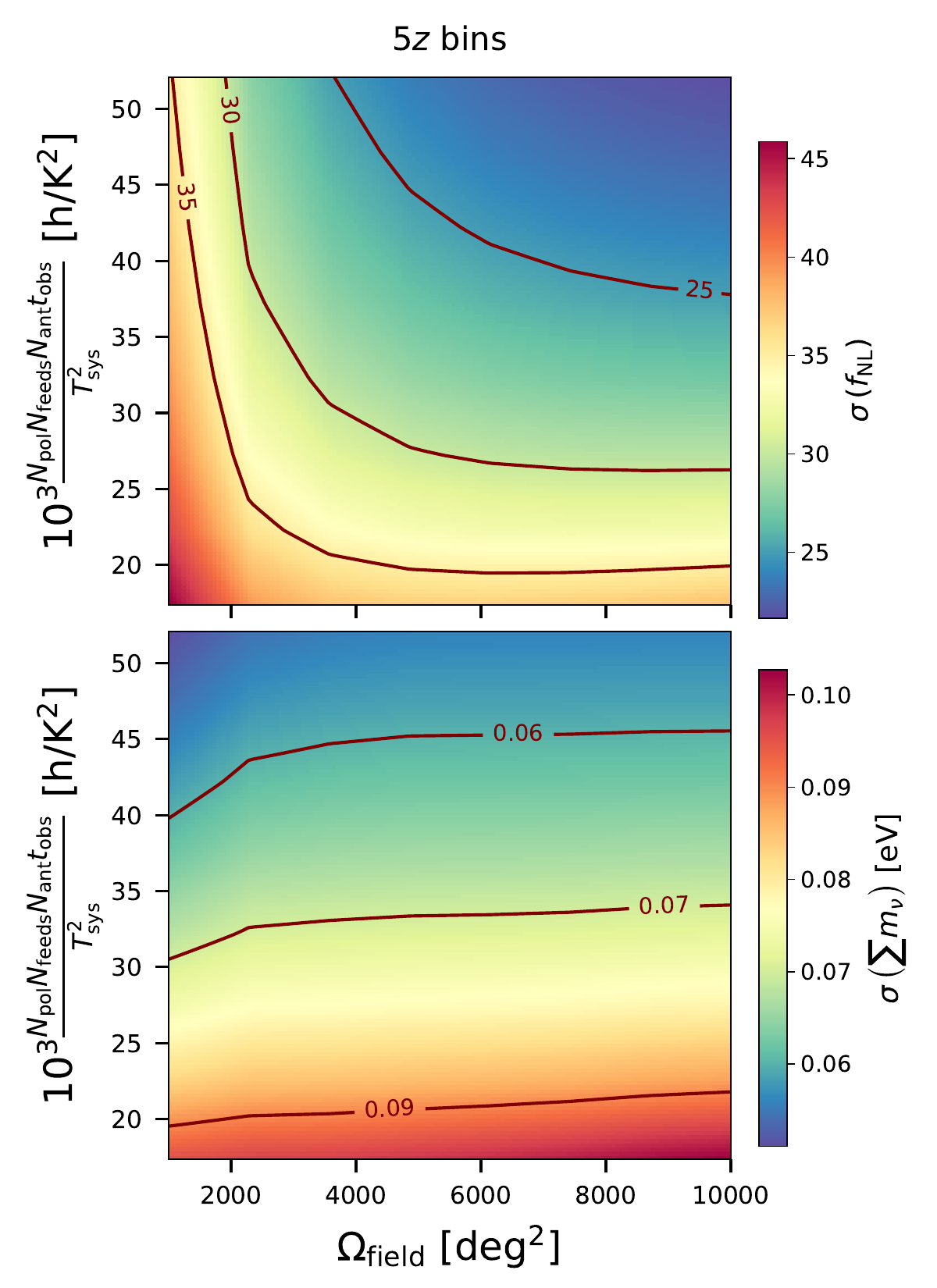}
\caption{Forecasted 68\% confidence-level marginalized  constraints on $f_{\rm NL}$ (top panels) and on $\sum m_\nu$ (bottom panels) for our general experiment using the monopole, quadrupole and hexadecapole, as function of $\Omega_{\rm field}$ and $N_{\rm pol}N_{\rm feeds}N_{\rm ant}t_{\rm obs}/T_{\rm sys}^2$, a combination of instrumental parameters which determines the measurement sensitivity. We compare results using a single redshift bin (Left panels) or five bins (Right panels). Note the change of scale in the color bars.}
\label{fig:opt_fNLmnu}
\end{figure*}

Fig.~\ref{fig:opt_fNLmnu} shows a comparison of the forecasted marginalized 68\% confidence-level constraints on $f_{\rm NL}$ and $\sum m_\nu$ from our generalized experiment using one and five redshift bins. It also shows  the dependence of the constraints on $\Omega_{\rm field}$ and $t_{\rm obs}N_{\rm feeds}/T_{\rm sys}^2$. While the former affects the volume probed (which determines $N_{\rm modes}$ and $W_{\rm vol}$), the latter only affects the amplitude of $P_n$. We restrict our investigation of the optimization of the survey to the variation of these parameters (for a comprehensive study of survey optimization to constrain $f_{\rm NL}$, see Ref.~\cite{Moradinezhad_Dizga_COfNL2}).

Having five redshift bins reduces $L_\parallel$ in each of them, which suppresses the power spectrum up to higher $k_\parallel^{\rm min}$ values, through $W_{\rm vol}$. This is the main reason that our forecasted errors on $f_{\rm NL}$ and $\sum m_\nu$ using a single redshift bin are approximately a factor of two better. For high values of $\Omega_{\rm field}$, sample variance is smaller than the instrumental noise, which is why the constraints improve so much when $t_{\rm obs}N_{\rm feeds}/T_{\rm sys}^2$ increases. The improvement is smaller for low $\Omega_{\rm field}$ since the sample variance contribution to the error dominates and reducing the instrumental noise does not significantly reduce the total covariance. Since the imprints of $f_{\rm NL}$ lie on large scales, this effect is more critical than for $\sum m_\nu$, which affects  a wider range of scales. Also, this effect is not very strong with five redshift bins, since the $\tilde{P}$ suppression limits the exploitation the very large line-of-sight scales.

\subsection{Further improvements}
\label{sec:further}

Besides the strategies explored above, the extraction of cosmological information from the LIM power spectrum can be further improved. One possibility would be to assign an arbitrary normalized weight, $w$, to each of the observed voxels in configuration space. Then,  $w(\vec{r})$ can be chosen to maximize the S/N of the measured IM power spectrum. With the introduction of these weights, a minimum-variance estimator can be derived, as proposed in Ref.~\citep{FKPweights} for galaxy surveys. An adaptation to LIM can be found in Ref.~\cite{Blake_Pkmod}.

Different redshift binning strategies or overlapping bins may be more optimal than the prescription used above. This will depend on the main objectives of the survey,  as argued in Section~\ref{sec:optimization}. However, if overlapping bins are to be used, they would not be independent anymore, and the corresponding covariance will have to be taken into account. 
Moreover, the actual optimal frequency band \change{(and corresponding redshift coverage)} of the experiment depends on the targeted parameters.  

Finally, in this work we have enforced the redshift bins to be narrow enough such that the redshift evolution of the large-scale brightness temperature fluctuations does not change significantly (except in Section~\ref{sec:optimization}). However, this can be incorporated into the analysis using redshift-weighting techniques~\citep{Zhu_redshiftweight}: redshift-dependent weights are chosen in order to minimize the projected error on the target cosmological or astrophysical parameter using a Fisher matrix forecast, so that the constraining power is maximized. These techniques can become especially useful when wide redshift bins are needed, e.g.\ for primordial non-Gaussianity, see Ref.~\cite{Mueller_fNL}.

\section{Conclusions}\label{sec:Conclusions}

LIM techniques have attracted substantial attention due to their untapped potential to constrain astrophysics and cosmology by probing huge volumes of the observable Universe which are beyond the reach of other methods. Nonetheless, the versatility of LIM can also be considered a nuisance, since cosmological information is  always intrinsically degenerate with astrophysics. These degeneracies, as well as the effect of subtle contributions to the observed LIM power spectrum and its covariance, as well as various assumptions entailed in the analyses, are too commonly ignored or only partially treated. 

The aim of this work has been to provide a comprehensive and general framework to optimally exploit the cosmological information encoded in the LIM power spectrum.
We have presented a reparameterization of it to identify and isolate the degeneracies between cosmology and astrophysics, including redshift-space distortions and the Alcock-Paczynski effect. Furthermore, we have introduced and advocated for the use of the multipole expansion of the LIM power spectrum. We also derived an accurate analytic covariance for the  multipoles, and discussed common errors in previous analyses.

Using a generalized experiment, and focusing on baryon acoustic oscillations and redshift-space distortions measurements, we showed that adding the hexadecapole to the monopole and quadrupole of the LIM power spectrum in the analysis returns a \NEWchange{$10\%-75\%$} improvement in the forecasted constraints. This is mainly due to the reduction of the correlation between the parameters, as the S/N of hexadecapole measurements is usually low. 
 However, redshift-space distortions as inferred from the LIM power spectrum are completely degenerate with $\langle T\rangle$. This degeneracy may be partially broken using mildly non-linear scales and perturbation theory, as proposed by Ref.~\cite{Castorina_21cmgrowth}. Other possible strategies include the combination with other cosmological probes~\cite{Obuljen_21cmcrossothers} or joint analyses of two-point and higher order statistics measurements~\cite{GilMarin_bispectrunNbody,Sarkar_21cmBispec}. 
 
 We also investigated different survey strategies depending on the target of the experiment. As an illustration, we compared forecasts for constraints on $\sum m_\nu$ and local primordial non-Gaussianity using one or multiple redshift bins over the same total volume, finding that in both cases using one redshift bin returns stronger constraints. We also explored the improvement of the constraints by decreasing the instrumental noise and increasing the size of the patch of the sky observed. While the constraints on $\sum m_\nu$ do not significantly improve when varying $\Omega_{\rm field}$, the size of the patch of the sky observed is critical to constrain primordial non-Gaussianity, since its signatures are dominant at large scales (however, increasing $\Omega_{\rm field}$ beyond $\sim 4000-6000$ deg$^2$ 
 does not improve significantly the constraints. 
   \change{Note that in some cases increasing $\Omega_{\rm field}$ even worsens the constraints. This is because the instrumental error, which grows due to the reduction of the observation time for each pointing, dominates the error budget on large scales.}
   
   Finally, we briefly discussed further improvements to the LIM power spectrum analyses to be studied in future works, especially to be applied to simulated and real observations, rather than theoretical work. In the Appendices, we provide an adaptation of our proposed framework to cross-power spectra between spectral lines or between LIM and galaxy surveys, and derive a close-to-optimal estimator for the true LIM power spectrum (i.e., without smoothing), and compare different approaches.
   
\change{The forecasted constraints in this work are calculated for a general non-specified experimental setup, as they are mostly intended for qualitative illustration and reference. The focus is on relative improvements and the performance of the methodology proposed; we leave forecasts for specific experiments to future work. Nonetheless, a comparison with the forecasted potential of future experiments targeting other cosmological probes can be useful. In the case of BAO measurements, our results show that LIM can be competitive with lower-redshift observations from spectroscopic galaxy surveys like DESI~\cite{desi}. Moreover, LIM BAO will be unique as it allows to probe higher redshifts than those reachable by galaxy surveys. The potential of LIM on its own to constrain $\sum m_\nu$ is promising, although likely not competitive with the combination of CMB probes and galaxy surveys~\cite{Brinckmann_mnuforecasts}. This should encourage the combination of CMB and LIM (with a more detailed modelling of the effects due to $\sum m_\nu$) to constrain $\sum m_\nu$. Finally, the LIM forecasted constraints on $f_{\rm NL}$ are inferior compared to the expected potential of other probes such as galaxy surveys (see e.g.,~Ref.~\cite{Bernal_EMU,redbook,desi}) or the kinetic Sunyaev-Zeldovic effect (see e.g., Ref.~\cite{Munchmeyer_kszfnl}). In any case, note that the potential of cross-correlating multiple lines has not been explored in this work, which is expected to significantly improve the results, as well as our knowledge about the astrophysics of the IGM (see e.g.,~\cite{Heneka_21cmLya}).}
 
\change{The formalism presented in this work aims to be applicable to any emission line at any redshift. However, there might be some cases where astrophysical complexities affect the spatial distribution of the intensity of the lines, mainly through extending the emission beyond the size of a dark-matter halo. In these  cases (e.g., radiative transfer for the Lyman-$\alpha$ line, reionization), the methodology as presented here might not be general enough and fail to marginalize over these effects. We leave the investigation of these scenarios to future work.}
 
 Though we have focused here on power spectrum analyses, line-intensity maps contain a significant amount of information beyond their power spectra.  Line-intensity fluctuations are generated by highly complex gas dynamics on sub-galactic scales, which gives rise to a significantly non-Gaussian intensity field.  \change{Although the power spectrum is also affected by the non-Gaussianities sourced by the inherent astrophysics of LIM and the non-linear collapse, alternative summary statistics, such as higher order correlations or the voxel intensity distribution~\cite{Breysse_VID}, are  needed in order to fully exhaust the information encoded in LIM observations. Ideally, these summary statistics will be measured and analyzed with the power spectrum in tandem, accounting for their correlation, as  first explored in Ref.~\citep{Ihle_VID-PS}.}
 
 The methodology developed in this work should find ample opportunities for implementation. As a compelling example, we demonstrate in a companion paper, Ref.~\citep{Bernal_IM_letter}, that LIM can be used to efficiently probe the expansion history of the universe up to extremely high redshifts ($z\lesssim9$), possibly weighing in on the growing Hubble tension~\cite{Freedman:2017yms,RiessH0_19,Wong:2019kwg} and suggested models to explain it (including models of evolving dark energy, exotic models of dark-matter, dark-matter--dark-energy interaction, modified gravity, etc., see e.g.\ Refs.~\cite{Karwal:2016vyq,Raveri_DEMG,Poulin_EDE,Stephon_H0dilaton,Lin:2019qug,Kreisch_selfnu,Yang:2019nhz,Archidiacono:2019wdp,Hooper:2019gtx,Vattis:2019efj,Pandey:2019plg,Pan:2019gop,Desmond:2019ygn,Agrawal:2019lmo,Vagnozzi:2019ezj}). It can also be used to constrain model-independent expansion histories of the Universe~\cite{BernalH0,Poulin_H0,Joudaki:2017zhq} at the few-percent level. Our formalism could also be adapted to measure the velocity-induced acoustic oscillations~\citep{Munoz_vao_meth}, recently proposed in Ref.~\cite{Munoz_vao} as a standard ruler at  cosmic dawn. 
 
We are eager for the next generation of LIM  observations to be available for cosmological analyses, and hope that the general framework reported in this manuscript can guide its precise and robust exploitation.

\acknowledgements
Funding for this work was partially provided by the Spanish MINECO under projects {AYA2014-58747-P} AEI/FEDER UE and MDM-2014-0369 of ICCUB (Unidad de Excelencia Maria de Maeztu).
JLB is supported by the Spanish MINECO under grant BES-2015-071307, co-funded by the ESF, and thanks Johns Hopkins University for hospitality during early stages of this work. HGM acknowledges the support from ”la Caixa” Foundation (ID 100010434) which code LCF/BQ/PI18/11630024”.

\appendix
\section{Formalism for cross-correlations}\label{sec:Cross}
\subsection{The IM cross-power spectrum}
Let us consider two generic brightness-temperature maps of two different spectral lines, denoted by $X$, $Y$, respectively. In this case, the Kaiser effect present in the RSD factor is different for each tracer. For instance, in $F^X_{\rm RSD}\propto  \left(1+f\mu^2/b_X\right)$, with $b_{X}$ being the luminosity-averaged bias for the line X. Then, the LIM cross-power spectra of the $X$ and $Y$ lines is given by:
\begin{equation}
\begin{split}
& P^{XY} = P^{XY}_{\rm clust}+ P^{XY}_{\rm shot}\,;\\ 
& P^{XY}_{\rm clust}= \langle T_X\rangle \langle T_Y\rangle b_Xb_Y F^X_{\rm RSD}F^{Y}_{\rm RSD}P_{\rm m};\, \\
& P^{XY}_{\rm shot}  = \left(\frac{c^3(1+z)^2}{8\pi k_B\nu^3H(z)}\right)^2\int dML_X(M)L_Y(M)\frac{{\rm d}n}{{\rm d}M}\,,
\end{split}
\label{eq:Pk_cross}
\end{equation}
where we have dropped the explicit notation regarding dependence on $k$, $\mu$ and $z$ (as will be done hereafter) for the sake of readability (See Refs. \cite{Wolz2017,Breysse2019} for a derivation). Following the same arguments as in Section~\ref{sec:ModellingPK}, we  express the cross-power spectrum between two different lines as:
\begin{equation}
\begin{split}
P^{XY} & =  \frac{P_{\rm m}/\sigma_8^2}{\left(1+0.5\left[ k\mu\sigma_{\rm FoG}  \right]^2\right)^2} \times\\
& \times\left(\tilde{T}_{XY}^{1/2} b_X\sigma_8  + \tilde{T}_{XY}^{1/2} f\sigma_8 \mu^2\right) \times \\
& \times\left(\tilde{T}_{XY}^{1/2} b_Y\sigma_8  + \tilde{T}_{XY}^{1/2} f\sigma_8 \mu^2\right) + P^{XY}_{\rm shot} ,
\end{split}
\label{eq:paramPk_cross}
\end{equation}
where $\tilde{T}_{XY}=\langle T_X\rangle\langle T_Y\rangle$. Since we have two different biases and two different brightness temperatures (one per line), the measurable combinations of parameters are: 
\begin{equation}
\begin{split}
\vec{\theta}_{XY} = \lbrace & \alpha_{\perp}, \alpha_\parallel, \tilde{T}_{XY}^{1/2} f\sigma_8, \tilde{T}_{XY}^{1/2} b_X\sigma_8, \tilde{T}_{XY}^{1/2} b_Y\sigma_8, \\
& \sigma_{\rm FoG}, P^{XY}_{\rm shot}, \vec{\varsigma} \rbrace\,.
\end{split}
\end{equation}

The  covariance of the LIM cross-power spectrum is also slightly  different. Considering, without loss of generality, the case in which each line is observed by a different experiment, the covariance per $\mu$ and $k$ bin of the cross-power spectra of two different lines  is (neglecting mode coupling) given by:
\begin{equation}
\tilde{\sigma}^2_{XY} = \frac{1}{2}\left(\frac{\tilde{P}_{XY}^2}{N_{\rm modes}} + \tilde{\sigma}_X\tilde{\sigma}_Y  \right),
\label{eq:sigma2_cross}
\end{equation}
where $\tilde{\sigma}_X$ and $\tilde{\sigma}_Y$ are the square root of the corresponding covariances in Eq.~\eqref{eq:sigma2}.

 Often, if the cross-power spectrum of two tracers can be measured,  the corresponding auto-power spectra can be as well. In this case, when comparing the model to  observations, all this information needs to be taken into account. In this case, Eqs.~\eqref{eq:SN}, \eqref{eq:chi2}, and~\eqref{eq:Fisher} are still correct, but both the data vector and the covariance matrix need to be changed. The former will include both auto- and cross-power spectra, hence $\vec{\Theta}$ would be the concatenation of $\vec{\Theta}_{XX}$, $\vec{\Theta}_{XY}$ and $\vec{\Theta}_{YY}$. In turn, the covariance matrix will be formed by four square blocks of the same size, where the diagonal blocks would be $\tilde{\mathcal{C}}_{XX}$ and $\tilde{\mathcal{C}}_{YY}$, and the off-diagonal blocks, $\tilde{\mathcal{C}}_{XY}$. 
 
\subsection{Cross-correlation with galaxy surveys}
LIM observations can also be cross-correlated with \change{the galaxy spatial distribution}. Denoting the galaxy catalog and related quantities with subscript/superscript $g$, we have an RSD factor $F^g_{\rm RSD}\propto \left(1+f\mu^2/b_g\right)$. In this case, the cross-power spectrum of galaxy number counts and LIM is:
\begin{equation}
\begin{split}
& P^{Xg} = P^{Xg}_{\rm clust}+ P^{Xg}_{\rm shot}\,;\\ 
& P^{Xg}_{\rm clust} = \langle T_X\rangle  b_Xb_g F^{X}_{\rm RSD} F^{g}_{\rm RSD}P_{\rm m}\,; \\
& P^{Xg}_{\rm shot}  = \frac{c^3(1+z)^2}{8\pi k_B\nu^3H(z)}\frac{\langle \rho_L^X\rangle_g}{n_g}\,,
\end{split}
\label{eq:Pk_crossgal}
\end{equation}
where $\langle \rho_L^X\rangle_g$ is the expected luminosity density  sourced only from the galaxies  belonging to the galaxy catalog used, and  $n_g$ is the number density of such galaxies \cite{Wolz2017,Breysse2019}. Note the difference in the shot noise terms in Eq.~\eqref{eq:Pk_crossgal} with respect to Eqs.~\eqref{eq:Pk} and~\eqref{eq:Pk_cross}. This is because contributions to $P_{\rm shot}^{Xg}$ come only from  the locations occupied by the galaxies targeted by the galaxy survey, and the shot noise between the galaxy distribution and the luminosity sourced elsewhere vanishes. We also assume that the shot noise of the galaxy power spectrum is Poissonian (i.e., $P_{\rm shot}^{gg} = 1/n_g$). Nonetheless, clustering and halo exclusion may introduce deviations from a Poissonian shot noise. This non-Poissonian contribution can change the amplitude of the shot noise and even induces a small scale dependence~\citep{Ginzburg_halomodel,Schmittfull_tracers}. 

Similarly to Eq.~\eqref{eq:paramPk_cross}, we prefer to express the cross-power spectrum of LIM and the galaxy number counts as:
\begin{equation}
\begin{split}
P^{Xg}& =  \frac{P_{\rm m}/\sigma_8^2}{\left(1+0.5\left[ k\mu\sigma_{\rm FoG}  \right]^2\right)^2} \times\\
& \times \left(\langle T_X\rangle^{1/2} b_X\sigma_8  + \langle T_X\rangle^{1/2} f\sigma_8 \mu^2\right) \times \\
& \times\left(\langle T_X\rangle^{1/2} b_g\sigma_8  + \langle T_X\rangle^{1/2} f\sigma_8 \mu^2\right) + P^{Xg}_{\rm shot} .
\end{split}
\label{eq:paramPk_crossgal}
\end{equation}
In this case, the parameter combinations measured would be:
\begin{equation}
\begin{split}
\vec{\theta}_{Xg} = \lbrace & \alpha_{\perp}, \alpha_\parallel, \langle T_{X}\rangle^{1/2} f\sigma_8, \langle T_X\rangle^{1/2} b_X\sigma_8,\\ 
& \langle T_X\rangle^{1/2} b_g\sigma_8, \sigma_{\rm FoG}, P^{Xg}_{\rm shot} , \vec{\varsigma} \rbrace\,.
\end{split}
\end{equation}
This parameterization already accounts for the possible variation of the amplitude of $P_{\rm shot}^{Xg}$ due to the non-Poissonian contributions mentioned above, since we have marginalized over $P^{Xg}_{\rm shot}$. Note that if the goal is to measure $\langle \rho_L^X\rangle_g$, the non-Poissonian contribution to the shot noise needs to be explicitly modelled. We neglect the potential scale dependence introduced, which is a good approximation at this stage.  As in the auto-spectrum case, one may be able to access some of the degenerate astrophysical information in this cross-spectrum through a map's one-point statistics, in this case by conditioning the LIM statistics on those of the galaxy catalog \cite{Breysse2019b}.

The covariance per $\mu$ and $k$ bin of the cross-power spectra of one line and galaxy number counts can be computed  (neglecting mode coupling) as:
\begin{equation}
\tilde{\sigma}^2_{Xg} = \frac{1}{2}\left(\frac{\tilde{P}_{Xg}^2}{N_{\rm modes}} + \tilde{\sigma}_X\frac{P_{gg}+\frac{1}{n_g}}{\sqrt{N_{\rm modes}}}  \right),
\label{eq:sigma2_crossgal}
\end{equation}
where $\left(P_{gg}+1/n_g\right)/\sqrt{N_{\rm modes}}$ is the square root of the covariance of the galaxy power spectrum (when Poissonian shot noise is assumed). 

\section{True vs. observed power spectra}\label{sec:Error}
In most of the literature, the effect of the instrument response, finite-volume surveyed and remaining residuals after foreground and interlopers removal is included in the error budget of the LIM power spectrum measurements, rather than incorporate it in the signal, as we do here. This implies that the  smearing of  $\delta T$ must be removed from the data (i.e., a deconvolution of the window function is required), and therefore the \textit{true} LIM power spectrum would be estimated in the analysis. 

\begin{figure}
\centering
\includegraphics[width=\linewidth]{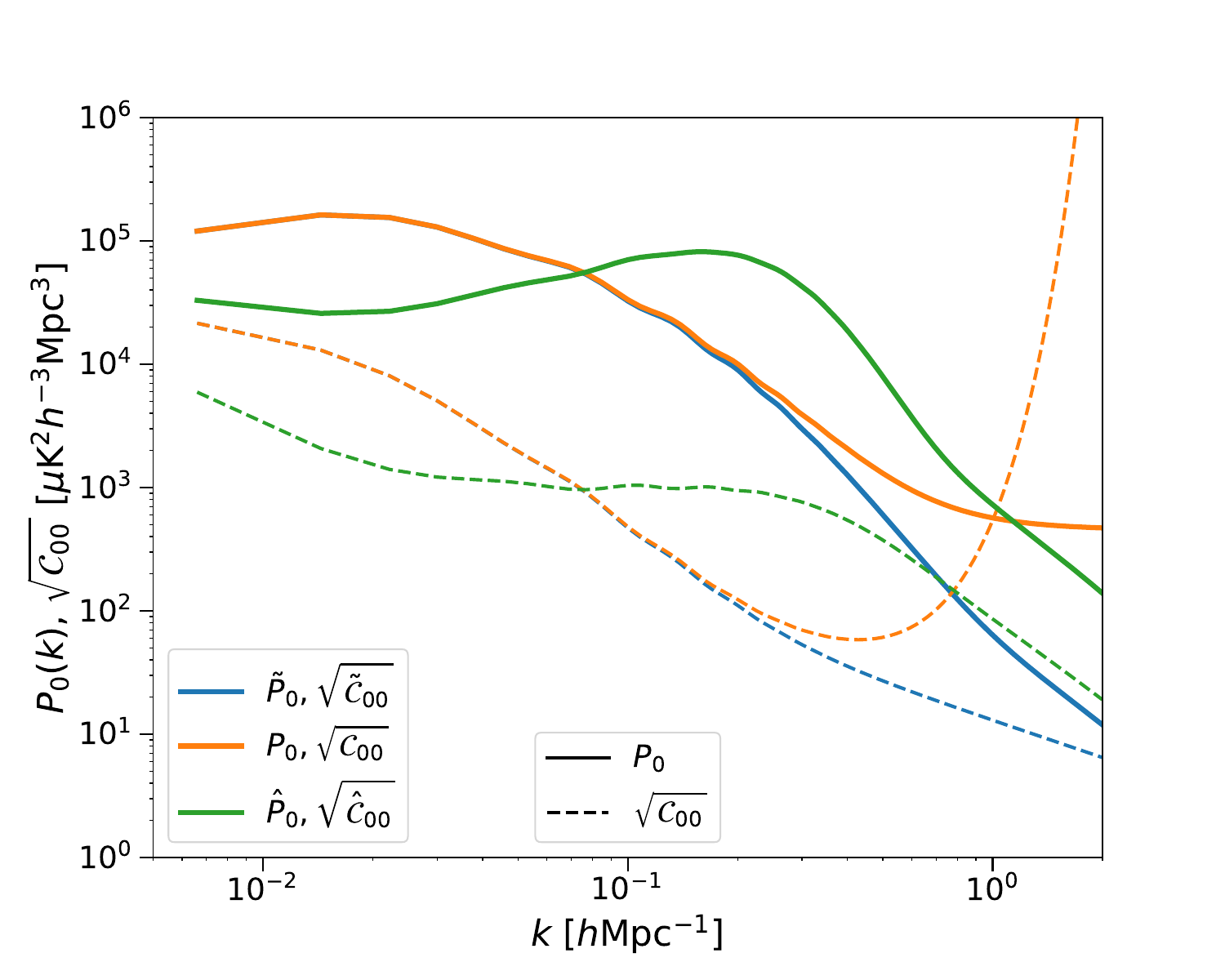}
\caption{Monopole of the LIM power spectrum (solid lines) and square root of its covariance (dashed lines) at $z=2.73$ for the generic experiment described in Section~\ref{sec:experiment}. Different colors denote different modelling of the LIM power spectrum and covariance: the observed one, $\tilde{P}$ (blue), the true one as in Eq.~\eqref{eq:sigma2prime}, $P$ (orange), and the true one with the weight scheme discussed in this Appendix, $\hat{P}$ (green). For illustration purposes, here we only consider $W_{\rm res}$ and ignore $W_{\rm vol}$.}
\label{fig:diffmods}
\end{figure}

The derivation of the observables, degeneracies and covariance is equivalent to the one presented in Sections~\ref{sec:ModellingPK} and~\ref{sec:covariance}, with two exceptions. On one hand, the summary statistic would be $P$ instead of $\tilde{P}$, which affects the power spectrum multipoles (Eq.~\eqref{eq:multipole_scale}), as shown in Fig.~\ref{fig:cosmovar}. On the other hand, since the noise power spectrum $P_n$, depicted in Eq.~\eqref{eq:Pn},  corresponds to the noise of the observed temperature fluctuations, the inverse of the window needs to be applied to $P_n$ (i.e., the same operation to obtain the \textit{true} power spectrum from the observations). In this case, the variance of $P$ per $k$ and $\mu$ bin (neglecting mode coupling) is:
\begin{equation}
\sigma^{2}(k,\mu) = \frac{1}{N_{\rm modes}(k,\mu)}\left(P(k,\mu) +\frac{P_n}{W(k,\mu)} \right)^2.
\label{eq:sigma2prime}
\end{equation}
Eq.~\eqref{eq:sigma2prime} has an immediate consequence on the S/N of the power spectrum multipoles. When computing $C_{\ell\ell^\prime}$ (following Eq.~\eqref{eq:covariance}, but using $\sigma^{2}$ instead of $\tilde{\sigma}^2$), the integral over $\mu$ tends to asymptote to infinity beyond the first resolution limit, i.e., $k\gtrsim \min\left(k^{\rm max}_\perp,k^{\rm max}_\parallel\right)$, due to the presence of $W$ in the denominator. The effect on $\mathcal{C}_{00}$ is shown in Fig.~\ref{fig:diffmods}, considering only $W_{\rm res}$. This behavior was unnoticed in previous work because either the anistropic power spectrum was directly used (rather than the multipoles), or due to an incorrect computation of the covariance. In many previous works, the thermal-noise contribution to the monopole error is given as proportional to $1/\int d\mu W(k,\mu)$, when in fact it should be proportional to $\int d\mu 1/W(k,\mu)$. Given the exponential behavior of $W$, $W_0^{-1}$ differs substantially from $(W^{-1})_0$, where the subindex $_0$ denotes the monopole.

However, an optimal estimator of the multipoles of $P$ should contain the same information that is contained in $\tilde{P}$, if $W$ is accurately modelled. In the following we discuss a next-to-optimal estimator of the multipoles of the \textit{true} power spectrum, $\hat{P}$. Let us consider that we could assign a weight, $w(\vec{k})$, to each mode of the brightness temperature perturbations in Fourier space, $\delta_kT$, such that the temperature fluctuations, $\mathcal{F}$, where $P = \langle\lvert \mathcal{F}\lvert^2\rangle$, becomes:
\begin{equation}
\mathcal{F} = \frac{1}{A^{1/2}}w\delta_kT\,,
\end{equation}
where $A = V_k^{-1}\int {\rm d}^3\vec{k}w^2$, $V_k$ is the volume in Fourier space, and all quantities depend on $k$ and $\mu$.  Hereinafter we will not show the explicit notation for the dependence for the sake of simplicity and readability. 

Applying the weights also to the noise power spectrum, we obtain $\hat{P}=w^2P/A$ and $\hat{P}_n=w^2P_n/A$. If we use these estimators for Eq.~\eqref{eq:sigma2prime}, the covariance of the multipoles of $\hat{P}$ is given by:
\begin{equation}
\begin{split}
\hat{\mathcal{C}}_{\ell\ell^\prime} = & \frac{(2\ell +1)(2\ell^\prime+1)}{2A^2N_{\rm modes}}\times
\\
& \times\int_{-1}^1{\rm d}\mu w^4\mathcal{P}\mathcal{L}_\ell(\mu)\mathcal{L}_{\ell^\prime}(\mu)\delta_{ij}^K,
\end{split}
\label{eq:covariance_estimator}
\end{equation}
where $\mathcal{P}=(P+P_n/W)^2$. In order to obtain the minimum variance using this framework, we impose the stability of the covariance under small changes in $w$.  Since most of the S/N comes from the monopole (see Fig.~\ref{fig:SNRk}), let us consider the minimization of the average of the covariance of the monopole in a shell in Fourier space. Imposing $w=w_0+\delta w$:
\begin{equation}
\begin{split}
&\frac{1}{V_k} \int_{V_k} {\rm d}^3\vec{k}\hat{\mathcal{C}}_{00}\propto \frac{\int {\rm d}^3\vec{k}w_0^4\left(1+\frac{\delta w}{w_0}\right)^4\mathcal{P}}{\left[\int {\rm d}^3\vec{k}w_0^2\left(1+\frac{\delta w}{w_0}\right)^2\right]^2}\approx \\
 &  \approx \frac{\int {\rm d}^3\vec{k} w_0^4\left(1+4\frac{\delta w}{w_0}\right)\mathcal{P}}{\left[\int {\rm d}^3\vec{k} w_0^2\left(1+2\frac{\delta w}{w_0}\right)\right]^2}\approx \\
 & \approx \frac{\int {\rm d}^3\vec{k} w_0^4\mathcal{P}}{\left[\int {\rm d}^3\vec{k} w_0^2\right]^2} \times \\
  & \times \left(1+4\frac{\int {\rm d}^3\vec{k} w_0^3\delta w\mathcal{P}}{\int {\rm d}^3\vec{k} w_0^4\mathcal{P}} -
  4\frac{\int {\rm d}^3\vec{k}w_0\delta w}{\int {\rm d}^3\vec{k} w_0^2}     \right),
 \end{split}
 \label{eq:weights_intermediate}
\end{equation}
where we have expanded over $\delta w/w_0$ up to second order. The stability condition of $\hat{\mathcal{C}}_{00}$ is fulfilled if:
\begin{equation}
w_0 = \mathcal{P}^{-1/2} = \left(P + \frac{P_n}{W}\right)^{-1}\,.
\label{eq:optw}
\end{equation} 
\change{One could even improve the weighting scheme with a matrix of the form $\mathcal{W}(\vec{k}_i,\vec{k}_j)$, but this would introduce artificial mode-coupling and complicate the measurement even more for the sake of a marginal gain. The exploration of this weighting scheme lies beyond the scope of the discussion in this appendix.}

We compare the three modellings of the LIM power spectrum and the covariances discussed in this appendix in Fig.~\ref{fig:diffmods}, only accounting for $W_{\rm res}$ and neglecting the effects of $W_{\rm vol}$ for the sake of clarity in this illustration. We show the monopole of the measured LIM power spectrum, $\tilde{P}$, of the true LIM power spectrum, $P$, and applying the weights of Eq.~\eqref{eq:optw} to the true LIM power spectrum, $\hat{P}$, as well as the square root of the corresponding covariances. As can be seen, using $P$ is very limited, since the S/N tends to zero at the first resolution limit. On the contrary, $\hat{P}$ shows a slightly more optimal behaviour than $\tilde{P}$. However, as discussed in Section~\ref{sec:MeasuredPK},  measuring $\delta_kT$ will be extremely difficult when the plane-parallel approximation breaks down. Therefore, we advocate for the use of $\tilde{P}$ and $\tilde{\mathcal{C}}_{\rm \ell\ell^\prime}$, since their measurement is more straightforward.

\bibliography{biblio}
\bibliographystyle{utcaps}

\end{document}